\documentclass[preprint,showpacs,preprintnumbers,amsmath,amssymb,floatfix,superscriptaddress]{revtex4}

\usepackage{graphicx}
\usepackage{dcolumn}
\usepackage{bm}
\usepackage{longtable}
\begin{document}

\title{Calculations of static dipole polarizabilities of alkali dimers. Prospects for alignment of ultracold molecules.}

\author{Johannes Deiglmayr}
  \affiliation{Physikalisches Institut, Universit\"at Freiburg, Hermann-Herder-Strasse 3, 79104 Freiburg, Germany.}
  \altaffiliation[Also at]{Laboratoire Aim\'e Cotton, CNRS, B\^at. 505, Univ Paris-Sud, 91405 Orsay Cedex, France}
\author{Mireille Aymar}
  \affiliation{Laboratoire Aim\'e Cotton, CNRS, B\^at. 505, Univ Paris-Sud, 91405 Orsay Cedex, France}
\author{Roland Wester}
  \affiliation{Physikalisches Institut, Universit\"at Freiburg, Hermann-Herder-Strasse 3, 79104 Freiburg, Germany.}
\author{Matthias Weidem\"uller}
  \affiliation{Physikalisches Institut, Universit\"at Freiburg, Hermann-Herder-Strasse 3, 79104 Freiburg, Germany.}
\author{Olivier Dulieu}
  \email{olivier.dulieu@lac.u-psud.fr}
  \affiliation{Laboratoire Aim\'e Cotton, CNRS, B\^at. 505, Univ Paris-Sud, 91405 Orsay Cedex, France}

\date{\today}

\begin{abstract}
The rapid development of experimental techniques to produce
ultracold alkali molecules opens the ways to manipulate them and to
control their dynamics using external electric fields. A
prerequisite quantity for such studies is the knowledge of their
static dipole polarizabilities. In this paper, we computed the
variations with internuclear distance and with vibrational index of
the static dipole polarizability components of all homonuclear
alkali dimers including Fr$_2$, and of all heteronuclear alkali
dimers involving Li to Cs, in their electronic ground state and in
their lowest triplet state. We use the same quantum chemistry
approach than in our work on dipole moments (M. Aymar and O. Dulieu,
J. Chem. Phys. {\bf 122}, 204302 (2005)), based on pseudopotentials
for atomic core representation, Gaussian basis sets, and effective
potentials for core polarization. Polarizabilities are extracted
from electronic energies using the finite-field method. For the
heaviest species Rb$_2$, Cs$_2$ and Fr$_2$ and for  all
heteronuclear alkali dimers, such results are presented for the
first time. The accuracy of our  results on atomic and molecular
static dipole polarizabilities is discussed by comparing our values
with the few available experimental data and elaborate calculations.
We found that for all alkali pairs, the parallel and perpendicular
components of the ground state polarizabilities at the equilibrium
distance $R_e$ scale as $(R_e)^3$, which can be related to a simple
electrostatic model of an ellipsoidal charge distribution. Prospects
for possible alignment and orientation effects with these molecules
in forthcoming experiments are discussed.

\end{abstract}


\maketitle

\section{Introduction}

The response of an atomic or molecular system to an external
electric field is driven in many situations by its static electric
polarizabilities which expresses the propensity of the electronic
structure to be affected by the field \cite{miller1977}. In this
respect, the static polarizability is sensitive to the details of
the electronic wave function of the system, and yields a constraint
for models aiming at evaluating it. The growing availability of
samples of cold and ultracold molecules \cite{doyle2004,dulieu2006}
open new routes to manipulate their motion in the laboratory frame
and control their dynamics using external electric fields.
Spectacular achievements concern polar molecules, i.e. molecules
with a permanent electric dipole moment: several species (CO, NO,
NH, NH$_3$,OH,H$_2$CO...) have been slowed down to kinetic energy
equivalent to a few millikelvins inside Stark decelerators
\cite{bethlem2003,bochinski2004,hudson2006}, and subsequently
trapped inside a storage ring (a "molecular synchrotron"
\cite{heiner2007}). Slow ND$_3$ \cite{junglen2004} and D$_2$O
\cite{rieger2006} molecules have been filtered out of a
Maxwell-Boltzmann distribution and guided through an electrostatic
quadrupole. In these arrangements, the response of the molecule to
the external field is dominated by its permanent dipole moment, even
if very high electric field values would be considered
\cite{gonzalez-ferez2004}. This effect is magnified when using a
Rydberg atom or molecule, which possess a permanent dipole moment
thousands times larger than the one for typical ground state
molecules \cite{vanhaecke2005}.

However, even non-polar species can be manipulated by strong
electric fields produced by far-off resonant laser fields through
the anisotropy of their static polarizability. Trapping in a quasi
electrostatic laserfield has been experimentally demonstrated with
cold Cs$_2$ molecules
\cite{takekoshi1998,wester2004,kraft2005,mark2007}, and cold
collisions between Cs atoms and trapped Cs$_2$ molecules have been
studied in such optical dipole traps \cite{zahzam2006,staanum2006}.
The deceleration and velocity bunching of a supersonic molecular
beam using traveling optical lattices has been proposed for the
iodine dimer \cite{barker2002}, and first observed for CO molecules
\cite{dong2004}. The interaction of molecules with strong polarized
laser fields is also well-known for yielding the possibility to
align the molecular axis along the electric field axis
\cite{friedrich1995,dion1999,sakai1999,larsen1999}, i.e. creating
pendular states. In such situations, the fast-oscillating electric
field averages its interaction with the permanent dipole moment of
polar molecules to zero, so that only the quadratic interaction
through the static polarizability persists. Recently, the
combination of a strong electrostatic field and a non-resonant laser
field has been proposed to enhance the orientation of polar
molecules \cite{friedrich1999}, and a first experimental evidence of
this effect has been reported on the HXeI complex
\cite{friedrich2003,nahler2003}. It is also worthwhile to mention
the recent proposal for controlling ultracold polar molecules in an
optical lattice combined with a suitable microwave field, relying on
the dynamic polarizability of the molecule \cite{kotochigova2006}.

Such developments could clearly benefit from the accurate knowledge of the structure and properties of the
concerned molecules. We started recently a new series accurate and systematic calculations of electronic
properties of all alkali pairs, both homonuclear and heteronuclear, which are up to now systems of choice for
ultracold molecules experiments. Our goal is to make available a complete treatment of these systems with the same
accuracy for all combinations. For instance, we computed the permanent electric dipole moments of all
heteronuclear species \cite{aymar2005} for which experimental data are still lacking for their ground state. We
displayed their variation with the interatomic distance $R$, as well as with the vibrational level. Most of these
results
were not previously available. We also investigated the transition dipole moments for numerous transitions of NaK,
NaRb, NaCs \cite{aymar2007} as only scattered results exist for the two latter species.  We extended our
calculations to the potential curves, and permanent and transition dipole moments of the francium diatomic compounds
Fr$_2$, RbFr and CsFr which were determined for the first time \cite{aymar2006}, and we investigated the
possibility of creating such diatomic compounds in an ultracold environment.

The present study deals with systematic calculations of static
dipole polarizabilities of alkali atoms and dimers. Static dipole
polarizabilities of ground state alkali atoms have been the subject
of numerous calculations as it can be seen in the review by Teachout
and Pack \cite{teachout1971}. Recently several systematic surveys of
alkali atomic polarizabilities
\cite{derevianko1999,safronova1999,lim1999,magnier2002} have been
published. On the experimental side, Molof {\it et al.} have
measured static dipole polarizabilities of all alkali atoms using a
$E-H$ gradient balance technique \cite{molof1974}, while other
authors performed a similar study using an electric-field deflection
method \cite{chamberlain1963,hall1974,miller1988}. Several
measurements have been obtained with other methods like Stark-shift
spectroscopy with lithium \cite{hunter1991}), atom interferometry
with sodium \cite{ekstrom1995} and lithium \cite{miffre2006}.

Experimental results for alkali pairs are much scarcer than for the
atoms. Most measurements have been achieved by deflecting a
molecular beam inside an inhomogeneous electric field, for Li$_2$,
Na$_2$ \cite{molof1974a,tarnovsky1993,antoine1999}, K$_2$, Rb$_2$,
Cs$_2$ \cite{molof1974a,tarnovsky1993}, and NaLi
\cite{graff1972,antoine1999}, NaK, KCs \cite{tarnovsky1993}, most
often at different temperatures of the gaseous sample. Knight {\it
et al} \cite{knight1985} reported a measurement for Na$_2$ and $K_2$
using a supersonic beam. As detailed later, various authors  have
computed the static dipole polarizabilities of homonuclear alkali
dimers while results on heteronuclear dimers are scarcer.

In this paper, we systematically calculate the static dipole
polarizabilities of all alkali atoms including francium, of all
homonuclear alkali dimers including Fr$_2$, and of heteronuclear
alkali dimers, involving Li to Cs, in their ground  and lowest
triplet states. We use the same quantum chemistry approach as in our
previous works \cite{aymar2005,aymar2006}. The parallel and
perpendicular components of the static dipole polarizabilities as
functions of the internuclear distance $R$ are obtained with the
finite-field method \cite{cohen1965}. We demonstrate that the
polarizabilities at the equilibrium distance $R_e$ scale as $R_e^3$,
so that all alkali dimers similarly behave as an ellipsoid charge
distribution in an electric field. We analyze the accuracy of our
calculations by comparing the average polarizabilities and their
anisotropy with available experimental data and elaborate
calculations. Finally, we discuss the possibility for the permanent
alignment and the orientation of such ultracold molecules in strong
optical and electrical fields. In the following we will use atomic
units except otherwise stated.


\section{Computational method}
\label{sec:details}

A detailed presentation of our approach for calculating electronic molecular structure of alkali dimers can be found in our
previous papers \cite{aymar2005,aymar2006}, and we only recall the main lines here. We set up an automated
procedure based on the CIPSI package (Configuration Interaction  by Perturbation of a multiconfiguration wave
function Selected Iteratively) \cite{huron1973}. The approach is based on the  $\ell$-dependent pseudopotentials
of Durand and Barthelat \cite{durand1974,durand1975} for atomic core representation, Gaussian basis sets, and
effective potentials to account for core polarization (CPP) \cite{muller1984,foucrault1992}.
Molecular  orbitals are determined by restricted  Hartree-Fock single-electron calculations, yielding the
potential curves for the relevant molecular cations. A full valence configuration interaction (CI) is then
performed for each involved molecular symmetry, providing potential curves and permanent and transition dipole
moments.

If the $z$ axis is chosen along the internuclear axis in a
molecule-fixed reference frame ($x$,$y$,$z$), they are two
independent components of the molecular polarizability tensor, i.e.,
the parallel component $\alpha_{\parallel}\equiv \alpha _{zz}$ and
the perpendicular one $\alpha_{\perp}\equiv \alpha_{xx} =
\alpha_{yy}$. Two related quantities are usually defined: the
average polarizability  $\overline{\alpha}=(\alpha _{zz}
+2\alpha_{\perp})/3$ and the polarizability anisotropy $\gamma =
\alpha _{\parallel}- \alpha _{\perp}$. Atomic and molecular static
dipole polarizabilities have been calculated using finite-field
method \cite{cohen1965} implemented in the quantum chemistry
approach above. For each molecular system we calculated the energies
at fixed $R$ for several values of the external electric field
(1$\times10^{-4}$ to 5$\times10^{-4}$~a.u. depending on the
molecule, with 1~a.u.$=5.142206281 \times 10^{11}$~V.m$^{-1}$) in
the perturbative regime. We extracted the polarizability from their
quadratic dependence against the electric field magnitude.

\begin{figure}[f]
\includegraphics[width=0.8\columnwidth]{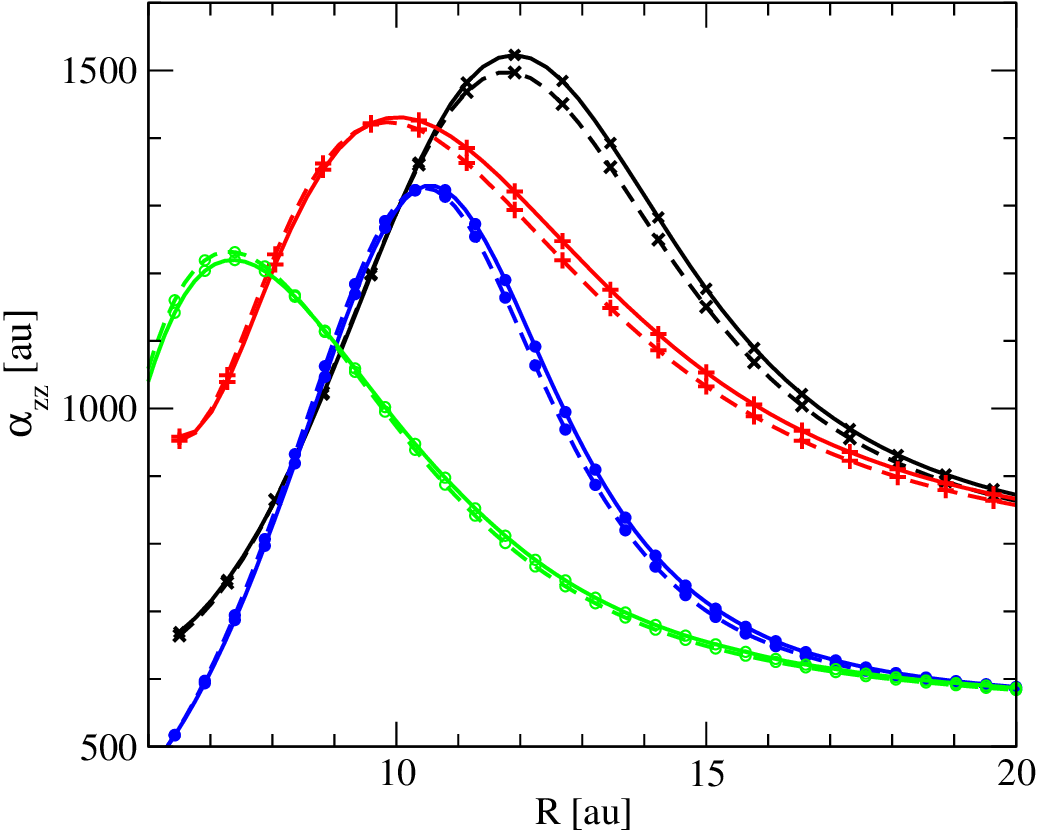}
\caption{\label{fig:basischeck} (Color on line) Parallel static
dipole polarizability (1~a.u.=0.1481847093~\AA$^3$) of the
$X^1\Sigma_g^+$ ground state and of the lowest $a^3\Sigma_u^+$
triplet state of Cs$_2$ (resp. crosses and plus symbols) and LiCs
(resp. closed circles and open circles) as computed with our quantum
chemistry approach using basis set A (full lines) or basis set B
(dashed lines) of Ref.~\cite{aymar2005} (of Ref.~\cite{aymar2006}
for Cs).}
\end{figure}

As for dipole moments, we checked the dependence of our results with
the size of the basis set, using both basis sets labeled A and B in
Ref.~\cite{aymar2005}. Figure \ref{fig:basischeck} illustrates this
influence on the $R$-dependence of $\alpha_{\parallel}$ for Cs$_2$
and LiCs ground state and lowest triplet state: the difference
between the two calculations never exceeds 1\%, which we will
consider as non-significant for our present purpose (see also the
discussion on atomic polarizabilities in the next section).

\section{Atomic static polarizabilities}

As usual in molecular calculations, we checked the quality of the
atomic representation yielded by our molecular basis with the
computation of the static dipole polarizabilities for all alkali
atoms (Li to Fr) with the finite-field method \cite{cohen1965}, and
compared them in table \ref{tab:atom} to various experimental
determinations and various recent high-precision calculations. We
first checked our values against the size of the basis set, using
both basis sets labeled A and B in Ref.~\cite{aymar2005}. The
differences in the obtained values never exceeded 1\% when using
basis A or B. For Li, K, Rb, and Cs atoms, our results for
$\overline{\alpha}$ agree at the 1\% level or better with the
experimental central values of Molof {\it et al.} \cite{molof1974}
obtained with a 10\% error bar. Our value for Na is also in
agreement at the same 1\% level with the improved experimental value
of Ekstrom {\it et al.} \cite{ekstrom1995}.

As shown in Table \ref{tab:atom}, all available elaborate
calculations agree with each other for the light species Li, Na, and
K. This is mainly due to the weakness of the relativistic effects in
these systems, whose magnitude can be evaluated through the work by
Kell\"o {\it et al.} \cite{kello1993}, who computed atomic
polarizabilities in the framework of a complete active space
self-consistent field (CASSCF) approach, where electron correlation
and relativistic effects are included as perturbations up to the
second-order. As expected, the relativistic contribution is hardly
noticeable for K, while it cannot be omitted for the heavier species
Rb, Cs, and Fr. This is probably the explanation of the discrepancy
found for these atoms with the ECP approach used in
Ref.~\cite{magnier2002}, so-called model potential approach, which
does not include explicitly such relativistic terms. In contrast,
the present ECP's include the effect of the mass-velocity and Darwin
terms, so that we obtain the same result than Ref.~\cite{kello1993}
for K, while this paper slightly overestimates the polarizabilities
for Rb, Cs, and Fr. Let us note that as already predicted in the
latter paper, the francium polarizability does not follow a
monotonic increase along the series of alkali atoms due to
relativistic effects: it is about 22\% smaller than the cesium one.
Let us note that we also found previously a manifestation of
relativistic effects in the permanent dipole moment of FrCs compared
to the RbCs one \cite{aymar2006}.

In contrast, our results are in very good agreement with those of
Safronova {\it et al.} \cite{safronova1999} and Derevianko {\it et
al.} \cite{derevianko1999}. These authors used the relativistic
single-double all-order (SD-AO) method combined with relativistic
random-phase approximation (RRPA). In the latter paper, the authors
claim that their theoretical values are in agreement but more
accurate than the experimental ones of Ref.~\cite{molof1974}, as
they introduced experimental values of energy levels and
high-precision experimental values for dipole matrix elements of the
principal transition in the computation of the polarizability. The
uncertainty on their values is then directly related to the
uncertainty of these experimental data. The difference with our
values remains within a 1\% to 2\% range. One of the largest
discrepancy is found for Cs, which could be due to the fact that we
do not use relativistic atomic orbitals in our calculations, in
contrast with francium for which we designed an averaged
relativistic core pseudopotential \cite{aymar2006}. Nevertheless,
our results confirm that in alkali species, relativistic effects
beyond mass-velocity and Darwin terms can be accounted for through
an averaged effective core potential, which would be certainly
relevant for Cs in further calculations. Using a relativistic
coupled-cluster approach, Lim {\it et al.} \cite{lim1999} also claim
that their values should be more accurate than the available
experimental ones. However, their values are systematically larger
(just like the relativistic values of Ref.~\cite{kello1993} than
ours and those of Ref.~\cite{derevianko1999} (except for Li and Na),
by 2\% to 3\% for K, by 2\% for Rb, and by 5\% for Cs and Fr.

To summarize, our atomic calculations seem to correctly account for
relativistic effects when they are noticeable, and represent a
satisfactory starting point for the computation of molecular
polarizabilities.

\begin{table}[h]
\center
\begin{tabular}{|c|c|c|c|c|c|c|c|} \hline
                                              &   &Li     &Na      &K       &Rb      &Cs       &Fr  \\ \hline
This work                                     &   &164.4  &164.1   &294.3   &318.1   &406.1    &316.6\\ \hline
Miffre {\it et al.} \cite{miffre2006}       &(E)&164.2(1)&&&&&              \\
Ekstrom {\it et al.} \cite{ekstrom1995}        &(E)&       &162.7(8)&&&&              \\
Molof {\it et al.} \cite{molof1974}             &(E)&164(3) &159(3)  &293(6)  &319(6)  &402(8) &\\
Hall and Zorn \cite{hall1974}                 &(E)&       &165(11) &305(22) &329(23) &427(31)&\\
\hline Derevianko {\it et al.} \cite{derevianko1999} &(T)&
&162.6(3)&290.2(8)&318.6(6)&399.9(1.9)&317.8(2.4)\\ \hline
Kell\"o {\it et al.} \cite{kello1993}a         &(T)&       &        &301.0   &410.3   &491.3     &552.0 \\
Kell\"o {\it et al.} \cite{kello1993}b         &(T)&       &        &295.6   &330.0   &413.7     &325.9 \\
Safronova {\it et al.} \cite{safronova1999}    &(T)&       &163.07  &290.10  &317.39  &399.8     &314.8 \\
Lim {\it et al.} \cite{lim1999}                &(T)&163.74 &164.89  &301.28  &324.24  &432.71    &330.70\\
Magnier and Aubert-Fr\'econ \cite{magnier2002}&(T)&164    &165     &302     &335     &434       &\\ \hline
\end{tabular}
\caption{Static dipolar polarizabilities (in atomic units) of alkali
atoms compared to available experimental (E) and recent theoretical
(T) works. Both non-relativistic (a) and relativistic (b) values
calculated by Kell\"o {\it et al.} \cite{kello1993} are displayed.}
\label{tab:atom}
\end{table}

\section{Static polarizability functions for homonuclear alkali dimers}
\label{sec:homopol}

\begin{figure}[f]
\includegraphics[width=0.8\columnwidth]{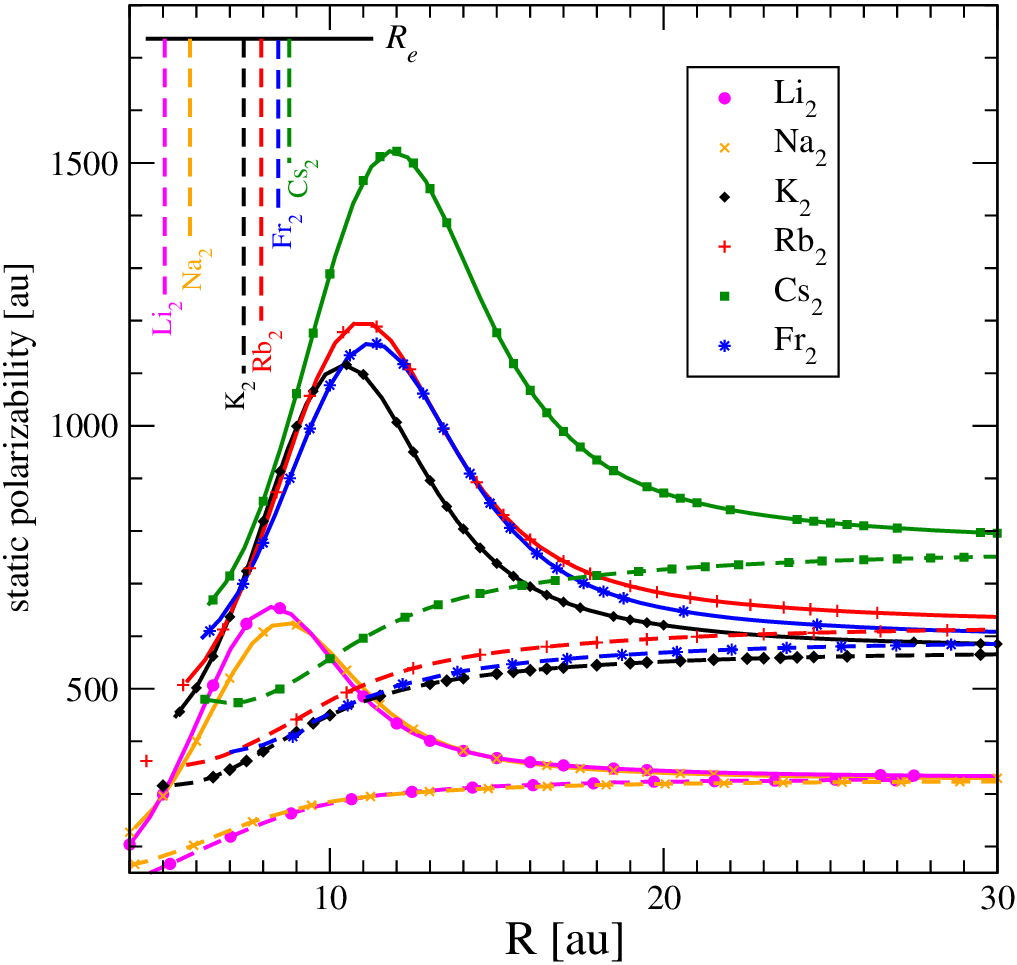}
\caption{\label{fig:homo_singlet} (Color on line) Parallel (full
lines) and perpendicular (dashed lines) static dipole polarizability
functions for the $X^1\Sigma_g^+$ ground state of homonuclear alkali
dimers, as computed in the present work. Experimental equilibrium
distances are also indicated for further discussion in the text.}
\end{figure}

\begin{figure}[f]
\includegraphics[width=0.8\columnwidth]{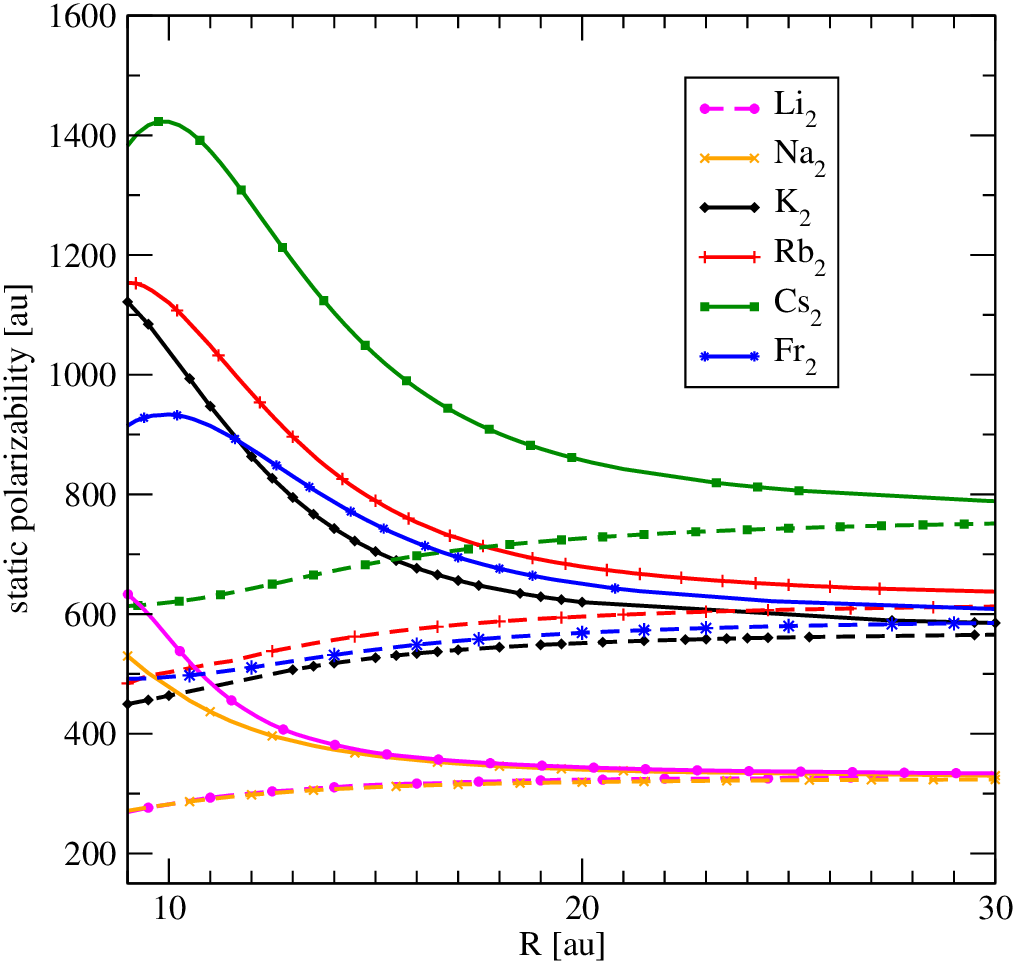}
\caption{\label{fig:homo_triplet} (Color on line) Parallel (full
lines) and perpendicular (dashed lines) static dipole polarizability
functions for the lowest $a^3\Sigma_u^+$ triplet state of
homonuclear alkali dimers, as computed in the present work. }
\end{figure}

As we will see in section \ref{sec:req}, most calculations of molecular static polarizabilities are restricted to
their value at the equilibrium distance. In Figures \ref{fig:homo_singlet} and \ref{fig:homo_triplet} we display
the variation of $\alpha_{\parallel}$ and $\alpha_{\perp}$ with the internuclear distance, respectively for the $X^1\Sigma_g^+$
ground state and for the lowest $a^3\Sigma_u^+$ triplet state of the homonuclear alkali dimers. The results for the heavy species Rb$_2$, Cs$_2$, and Fr$_2$ are presented for the first time, while to our knowledge, only one other theoretical determination was available for the ground
state of the light species Li$_2$, Na$_2$, and K$_2$ \cite{muller1986}, and for the lowest triplet state of Li$_2$, Na$_2$ \cite{rerat2003}, and K$_2$ \cite{merawa2003}.

The $R$-variation of $\alpha _{\parallel}$ and  $\alpha_{\perp}$ are
similar for all systems, and their magnitude increases with
increasing mass, as expected when the electronic clouds become
larger and larger. As seen above for the francium atom, the francium
dimer polarizability is predicted with the same magnitude than the
Rb$_2$ one, which is again a manifestation of the contraction of
electronic orbitals due to relativistic effects. The parallel
polarizabilities for the ground state exhibits a maximum at a
distance around 1.3 to 1.5 times the equilibrium distance $R_e$ of
the dimers. In contrast the perpendicular components always have a
smaller magnitude than $\alpha_{\parallel}$, and monotonically
increase towards the asymptotic limit. At large distances, the
polarizability components converge toward twice the atomic values
$\alpha_{at}$, and as already noted by M\'erawa and R\'erat
\cite{merawa2003}, the parallel component varies two times faster
than the perpendicular one and with an opposite variation. This is a
well-known result of the asymptotic atom-atom picture, which yields
$\alpha_{\parallel}=2\alpha_{at}+4\alpha^2/R^3$ and
$\alpha_{\perp}=2\alpha_{at}-2\alpha^2/R^3$.

The magnitude and variations of the polarizabilities for the triplet
state are very similar to those of the ground state. However, the
maximum in the parallel component occurs in the region of the
repulsive wall of the triplet state, so that it monotonically
decreases with increasing $R$ over the range of the triplet
potential well.

\begin{figure}[f]
\includegraphics[width=0.8\columnwidth]{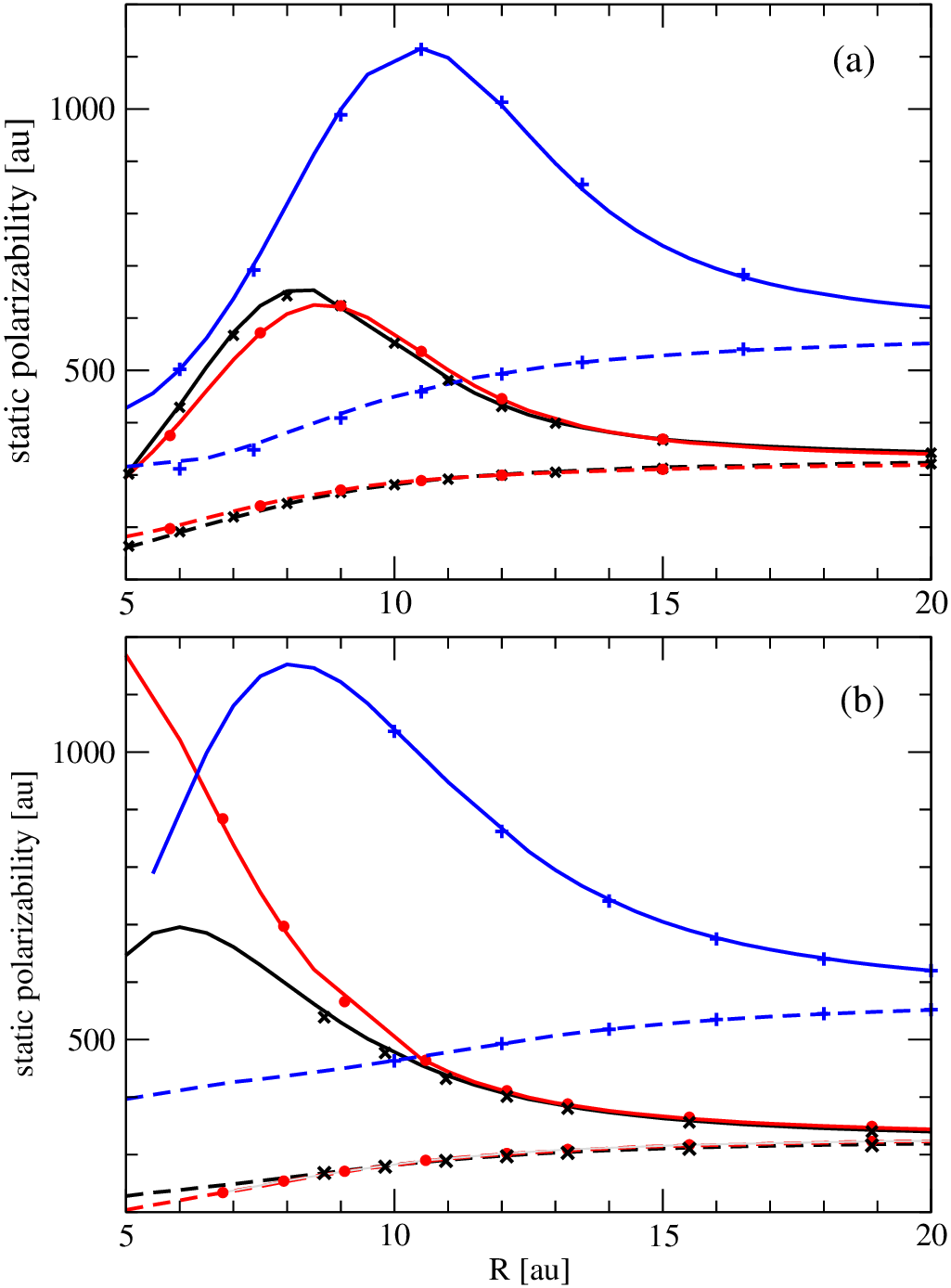}
\caption{\label{fig:homo_other} (Color on line) Comparison of the
present $\alpha_{\parallel}(R)$ (full lines) and
$\alpha_{\perp}(R)$(dashed lines) functions of Li$_2$, Na$_2$ and
K$_2$ (a) with those of M\"uller and Meyer \cite{muller1986} for the
ground state (symbols), and (b) with those of R\'erat and
Bussery-Honvault \cite{rerat2003,merawa2003} for the lowest triplet
state (symbols). Crosses, closed circles, and plus signs hold for
Li$_2$, Na$_2$, and K$_2$ respectively.}
\end{figure}

In Figure \ref{fig:homo_other}a, we compare the present
$R$-dependent polarizabilities to those computed with the
finite-field method \cite{cohen1965} by M\"uller and Meyer
\cite{muller1986} for the Li$_2$, Na$_2$, and K$_2$ ground state,
obtained by all-electron {\it ab initio} calculations which include
complete valence self-consistent approach, configuration
interaction, and core-polarization potentials to account for
core-valence correlation \cite{muller1984}. Our functions for the
Li$_2$, Na$_2$ and K$_2$ lowest triplet states are compared in
Figure \ref{fig:homo_other}b to the determination of
Refs.~\cite{rerat2003,merawa2003}, based on an asymptotic expression
for static polarizabilities~\cite{heijmen1996}. In both cases, the
agreement among all these quite different approaches is very good,
which is a convincing argument to assess the accuracy of our results
for the heavier species.

\section{Static polarizabilities functions for heteronuclear alkali dimers}
\label{sec:heteropol}

The $R$-dependent static polarizabilities for all heteronuclear
alkali pairs but NaLi (see Ref.~\cite{antoine1999,merawa2003a}) are
computed here for the first time. Indeed, the main response of such
systems to external electric fields is expected from their permanent
dipole moments, on which most theoretical studies concentrated in
the past (see for instance Ref.~\cite{aymar2005}). However, dipolar
traps for ultracold atomic or molecular systems rely on the
magnitude of static polarizabilities, while we already mentioned
above the proposal by Friedrich and Herschbach \cite{friedrich1999}
to manipulate polar or non-polar molecules with a combination of
strong laser field and a weak static field, to enhance their
orientation along the electric-field axis.

\begin{figure}[f]
\includegraphics[width=0.8\columnwidth]{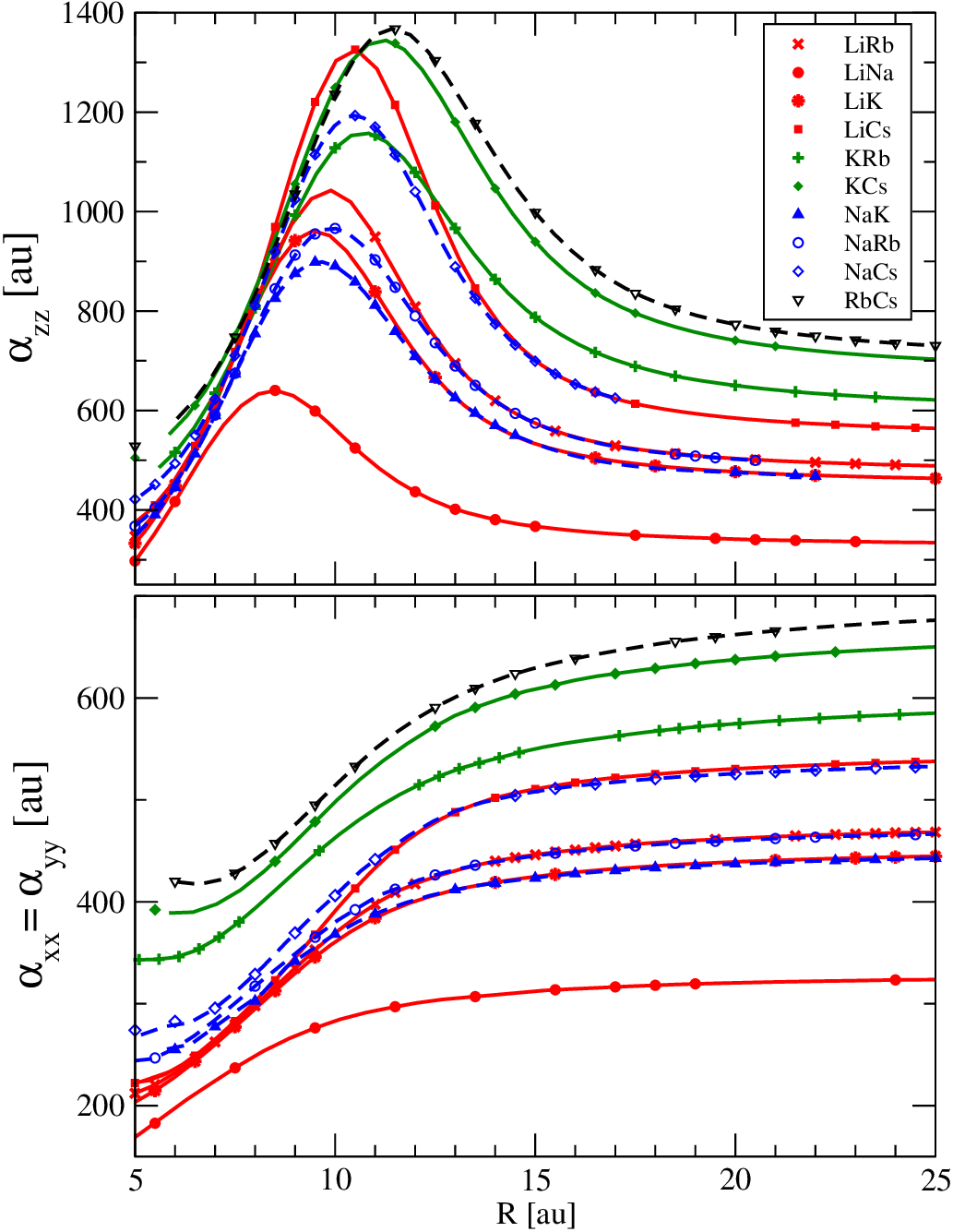}
\caption{\label{fig:hetero_singlet} (Color on line) Parallel (upper
panel) and perpendicular (lower panel) static dipole polarizability
functions for the $X^1\Sigma^+$ ground state of heteronuclear alkali
dimers, as computed in the present work. }
\end{figure}

\begin{figure}[f]
\includegraphics[width=0.8\columnwidth]{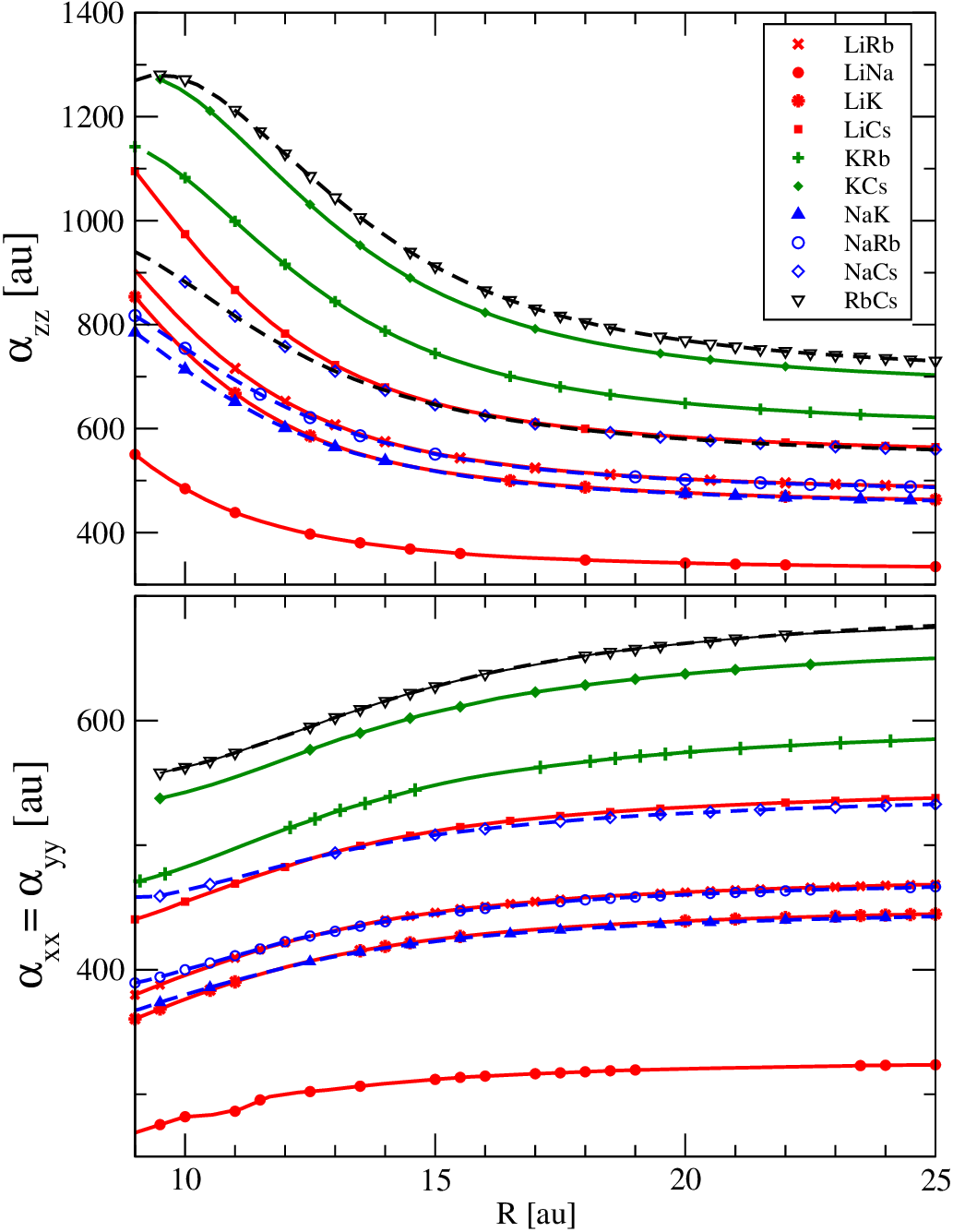}
\caption{\label{fig:hetero_triplet} (Color on line) Parallel (upper
panel) and perpendicular (lower panel) static dipole polarizability
functions for the lowest $a^3\Sigma^+$ triplet state of
heteronuclear alkali dimers, as computed in the present work. }
\end{figure}

As it can be seen in Figures \ref{fig:hetero_singlet} and
\ref{fig:hetero_triplet}, the $R$-variation of the static
polarizability components are similar to those for the homonuclear
species: the same shift by a factor of 1.3 to 1.5 of the maxima of
the parallel component compared to the equilibrium distance is
visible for the ground state, while the maxima in the triplet state
lies in the range of the repulsive wall of the triplet state. As
expected, the pairs containing heavy species are more polarizable
than those including light species. The RbCs ground state is then
found with the highest parallel polarizability around the
equilibrium distance, with the same magnitude than the one of KCs
and LiCs. In contrast, for all other cases the hierarchy among the
polarizabilities is governed by the ordering of the sum of atomic
polarizabilities. As the atomic polarizability of lithium and sodium
are almost equal and the smallest ones of the alkali atoms, pairs
involving Li or Na, and another atom among K, Rb, or Cs, exhibit a
polarizability magnitude quite close to each other, i.e., dominated
by the polarizability of the heavy species inside the concerned
pairs. We note that for each heteronuclear AB molecule the maximum
value of $\alpha _{\parallel}$ is given with a good approximation by
the averaged values of the corresponding homonuclear A$_2$ and B$_2$
species.

\section{Static polarizabilities at the molecular equilibrium distance}
\label{sec:req}

\begin{figure}[f]
\includegraphics[width=0.8\columnwidth]{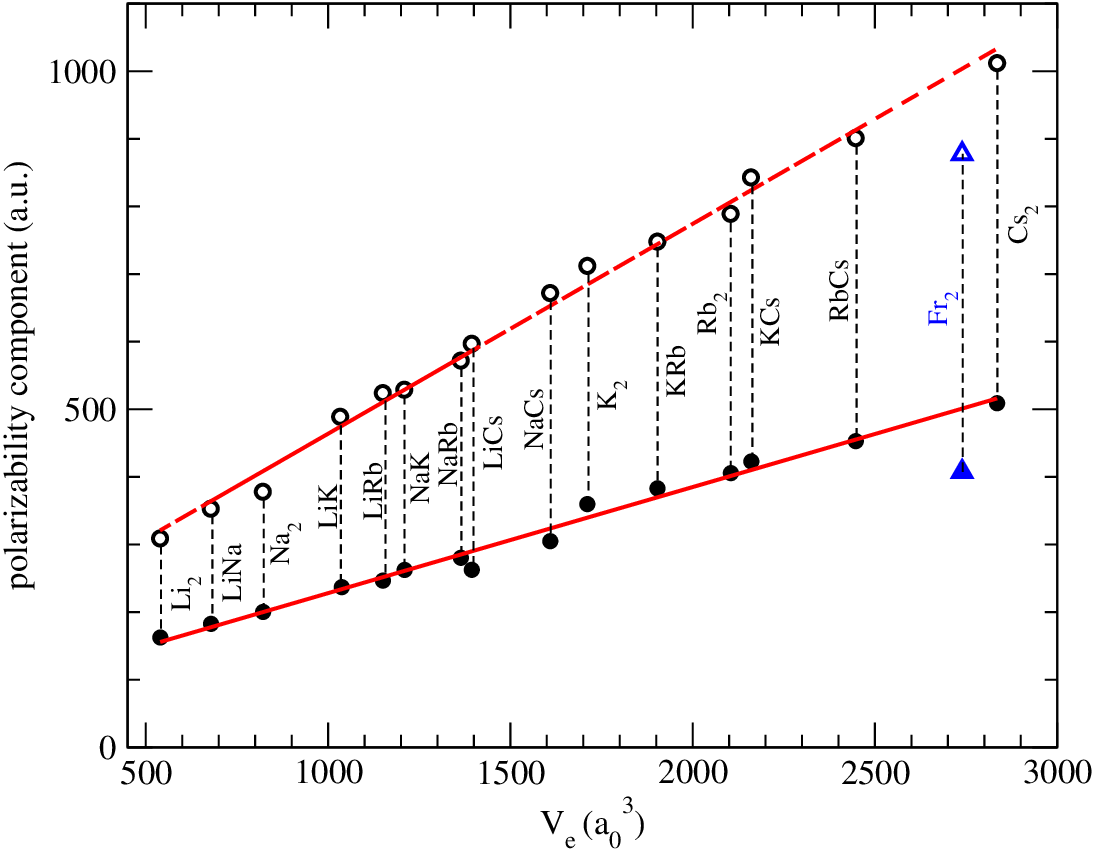}
\caption{\label{fig:pola_req} (Color on line) Parallel (full
circles) and perpendicular (open circles) static dipole
polarizabilities as functions of $V_e=4\pi R_e^3/3$, where $R_e$ is
the equilibrium distance of the ground state of every alkali pair.
The straight lines show a linear fit of this variation excluding
Fr$_2$ values, corresponding to the formula (in atomic units):
$\alpha_{\parallel}=0.31 V_e+153.5$ (dashed line) and
$\alpha_{\perp}=0.16 V_e+71.2$ (full line).}
\end{figure}

In contrast with previous sections, much more work has been devoted
to the computation of the static polarizabilities at the equilibrium
distance of the alkali pairs, as it is the value which could be
accessible for instance from deflection experiments of molecules in
the lowest vibrational level of their ground state. In this respect,
there is an interesting output of the present systematic study
performed with comparable accuracy for all alkali pairs. It is
well-known from classical electrostatics that the polarizability of
a charge distribution is proportional to its volume. If we plot the
ground state polarizabilities at the equilibrium distance $R_e$ of
homonuclear alkali and heteronuclear alkali dimers as a function of
$(R_e)^3$ (Figure \ref{fig:pola_req}), it is striking that all
species are well aligned. A linear fit shows that the parallel
component varies two times faster than the orthogonal component. As
expected, the francium dimer deviates from this phenomenological
law, due to its strong relativistic character.

Many authors already worked at designing models to relate the
polarizability of atomic and molecular systems to an effective
volume. Politzer {\it et al.} \cite{politzer2002} have shown that
atomic polarizabilities are proportional to an atomic volume derived
from the mean radii of Hartree-Fock  outer-shell orbitals. For
different types of molecules, Laidig and Bader \cite{laidig1990} and
Brinck {\it et al.} \cite{brinck1993} have also found a
proportionality between polarizabilities and molecular volume
calculated within a self-consistent field approach. Specific studies
performed on hydrocarbons molecules \cite{gough1989} or on sodium
clusters \cite{chandrakumar2004} have found a linear correlation
between the molecular polarizability and a characteristic volume.
Other models have been elaborated in different contexts, based on
the electrostatics of an ellipsoidal charge distribution like the
jellium model \cite{apell2002}. For instance, Ambj\"ornsson and
Apell \cite{ambjornsson2003} investigated the drift of ellipsoidal
polarizable particles through a viscous fluid induced by an electric
field gradient. Kornyushin \cite{Kornyushin2004} proposed a general
model for dipole plasma oscillations in an ellipsoidal sample.
Following the latter author, one can derive an expression for the
polarizability components $\alpha_b$ and $\alpha_a$ along the
principal axis of a symmetrical ellipsoid with a longitudinal radius
$b$ and a transverse radius $a$:

\begin{equation}
\alpha{_\perp}\equiv
\alpha_a=(2\frac{a^2}{b^2}+1)\frac{V}{4\pi}\hspace{1cm}\alpha{_\parallel}\equiv
\alpha_b=(\frac{b^2}{a^2}+2)\frac{V}{4\pi} \label{eq:ellipse}
\end{equation}

where $V=\frac{4\pi}{3}ab^2$ is the volume of the ellipsoid. One immediately sees on this expression than the ratio of 2 between both components is obtained if $b=\sqrt{2}a$. It is not straightforward however to deduce an effective radius for each alkali pair from the fitting formula reported in Figure \ref{fig:pola_req}, as the lines do not go through the origin.

Various theoretical methods have been used to compute the two
components of the static dipole polarizabilities of alkali dimers,
and we recollected several of them in the following tables to
compare with our present values. Generally, the extraction of static
polarizabilities relies on two steps: the calculation of electronic
structure, for which almost all available modern methods have been
considered, and the extraction of the polarizability values
themselves through a perturbative approach in most cases.

Bishop and Pouchan \cite{bishop1984} have employed a pseudopotential
treatment for core electrons and  included core-valence correlation
through a configuration interaction (CI) treatment. In a next paper,
Bishop {\it et al.} \cite{bishop1985a} have performed all-electron
self-consistent field (SCF) and CI calculations for electronic
structure (just like in the work of M\"uller and Meyer
\cite{muller1986}), and extracted the polarizability values via a
charge perturbation approach, deduced from the energies of the
system perturbed by a charge -1 located at 25$a_0$
($a_0=$0.0529177~nm) from the center of the molecule. Spelsberg {\it
et al.} extended the work of Ref.~\cite{muller1986} by computing
static polarizabilities for Li$_2$, Na$_2$, and K$_2$ at different
levels of approximation in their full CI treatment
\cite{spelsberg1993}. Polarizabilities of Na$_2$ and Na$_3$ have
been determined by Moullet {\it et al.} \cite{moullet1989} using
pseudopotential local-spin density (LSD) calculations. The CIPSI
approach for molecular electronic structure has been combined with
the time-dependent gauge invariant method (TDGI) by M\'erawa and
Dargelos \cite{merawa1998a} to calculate static polarizabilities,
dynamic polarizabilities and Van der Waals coefficients Li, Na,
Li$_2$ Na$_2$ and NaLi, while Antoine {\it et al.}
\cite{antoine1999} used Density Functional Theory (DFT) and CI for
the same purpose. An SCF and many-body perturbation theory (MBPT)
treatment, combined with finite-field method has been used by
Maroulis \cite{maroulis1988,maroulis2004} for Li$_2$ and Na$_2$
molecules. Urban and Sadlej \cite{urban1995} combined the MBPT and
coupled-cluster (CC) theories at different levels of approximation
and finite-field perturbation to determine static polarizabilities
of homonuclear and heteronuclear dimers not involving the Cs atom.
More recently, newly adjusted energy-consistent
nine-valence-electron scalar pseudopotentials including effective
CPP \cite{muller1984} have been developed by Lim {\it et al.}
\cite{lim2005} to investigate the properties of alkali dimers from
K$_2$ to Fr$_2$. Calculation of spectroscopic properties including
static dipole polarizabilities have been done using various models
based on CC and DFT theories, combined with finite-field method.

Table \ref{tab:homo_reqX} compares our values for the two components
$\alpha _{\parallel}$ and $\alpha_{\perp}$, and for the average
polarizability $\overline{\alpha}$ and its anisotropy $\gamma$ taken
either at the experimental equilibrium distance (when available) or
at the computed one \cite{aymar2005} of each pair, with the previous
theoretical determinations quoted above, in the case of the ground
state homonuclear alkali dimers. Due to the amount of papers already
published on Li$_2$, we quote only the most recent data, and the
interested reader could find a more complete compilation of older
works in Ref.~\cite{muller1986}. The above-quoted authors often
reported several polarizability values obtained within various
approximations, and we only display in Table \ref{tab:homo_reqX} the
value corresponding to their most elaborate model.

\begin{longtable}{|c|c|c|c|c|c|c|} \hline
                                      &&$R_e$ $(a_0)$&$\alpha_{\parallel}$&$\alpha_{\perp}$&$\overline{\alpha}$&$\gamma$ \\ \hline
Li$_2$($X^1\Sigma_g^+$)
&This work                              &5.051 \cite{barakat1986}&305.2&162.4&210.0&142.8 \\
&Bishop and Pouchan \cite{bishop1984}   &5.051 \cite{barakat1986}&357  &140  &213.3&217\\
&Bishop {\it et al.} \cite{bishop1985a}   &5.051 \cite{barakat1986}&324  &173  &223.0&151 \\
&M\"uller and Meyer \cite{muller1986}   &5.051 \cite{barakat1986}&301.8&169.9&213.8&131.9 \\
&Maroulis \cite{maroulis1988}           &5.051 \cite{barakat1986}&292  &170  &210.6&122\\
&Urban and Sadlej \cite{urban1995}      &5.051 \cite{barakat1986}&309.7&169.2&216.0&140.5\\
&M\'erawa and Dargelos \cite{merawa1998a}&5.051 \cite{barakat1986}&310.4&169.2  &216.1&141.4\\
&M\'erawa and R\'erat  \cite{merawa2001}&5.024                   &303  &160  &208  &143\\
&Antoine {\it et al.} \cite{antoine1999} &5.12                   &303.8&171.4&215.5&132.4\\ \hline
Na$_2$($X^1\Sigma_g^+$)
&This work                                 &5.818 \cite{babaky1988}  &378.5&199.6&259.2&178.9 \\
&M\"uller and Meyer \cite{muller1986}      &5.818 \cite{babaky1988}  &375.5&197.2&256.6&178.2 \\
&Moullet {\it et al.} \cite{moullet1989}    &5.818 \cite{babaky1988}  &318.5&199.73&259.1&179.3 \\
&Maroulis \cite{maroulis2004}              &5.818 \cite{babaky1988}  &377.7&206.6&263.3&171.7 \\
&Urban and Sadlej \cite{urban1995}         &5.818 \cite{babaky1988}  &386.9&209.7&268.7&177.2 \\
&M\'erawa and Dargelos \cite{merawa1998a}   &5.818 \cite{babaky1988}  &375.3&208.2&263.9&167.1 \\
&Antoine {\it et al.} \cite{antoine1999}    &5.84  \cite{babaky1988}&360.4&207.8&258.7&152.5\\ \hline
K$_2$($X^1\Sigma_g^+$)
&This work                                 &7.416 \cite{amiot1991}    &708.2&359.6&475.8&348.1 \\
&M\"uller and Meyer \cite{muller1986}      &7.379 \cite{muller1984}   &691.8&348.0&462.6&343.9 \\
&Urban and Sadlej \cite{urban1995}         &7.379 \cite{muller1984}   &753.6&376.2&502.0&377.4\\
&Lim {\it et al.} \cite{lim2005}            & 7.408 \cite{engelke1984a}&712.2&374.0&486.7&337.2\\
&Spelsberg {\it et al.}\cite{spelsberg1993} &7.379 \cite{muller1984}&677.8&363.3&468.1&314.5\\ \hline
Rb$_2$($X^1\Sigma_g^+$)
&This work                          &7.956 \cite{seto2000}  &789.7&405.5&533.5&348.2 \\
&Urban and Sadlej \cite{urban1995}  &8.1225 \cite{urban1995} &916.1&445.4&602.3&470.7 \\
&Lim {\it et al.}  \cite{lim2005}    &7.90
\cite{caldwell1980}&815.2&419.9&551.6&395.3 \\ \hline
Cs$_2$($X^1\Sigma_g^+$)
&This work                      &8.78 \cite{amiot2002a}&1012.2&509.0&676.7&503.2 \\
&Lim {\it et al.}  \cite{lim2005}&8.77 &1073.7&536.9&715.8&536.8 \\
\hline Fr$_2$($X^1\Sigma_g^+$)
&This work                      &8.45 \cite{aymar2006}&844.8&405.9&552.2&438.9 \\
&This work                      &8.6795 \cite{lim2005}&881  &408  &565.5&473 \\
&Lim {\it et al.}  \cite{lim2005}&8.6795
\cite{lim2005}&848.2&408.4&603  &367.8 \\ \hline

\caption {Present polarizability values (in a.u.) for the ground
state of homonuclear alkali dimers, taken at the experimental
equilibrium distance (except for Fr$_2$), and compared to available
theoretical works. For the latter, the distance at which the
polarizabilities are calculated is also displayed, when available.}
\label{tab:homo_reqX}
\end{longtable}

 One immediately
sees in Table \ref{tab:homo_reqX} the broad dispersion of the
reported values, which can be understood by looking again at Figure
\ref{fig:homo_singlet}. The equilibrium distance is located in the
steep part of the polarizability functions, so that any small
difference between the electronic wave functions yielded by the
various methods will result in a large variation of the
polarizability components, which will be even enhanced for the
average polarizability and the anisotropy. For the lighter species
Li$_2$ and Na$_2$, our values for $\alpha _{\parallel}$ and
$\alpha_{\perp}$ agree well with those of M\"uller and Meyer
\cite{muller1986} and of Urban and Sadlej \cite{urban1995}, as well
as with the other recent determinations. However, for K$_2$, the
value of Ref.~\cite{urban1995} seems to be overestimated by about
7\%, and by about 16\% for Rb$_2$. The values of Lim {\it et al.}
\cite{lim2005} are also slightly larger than ours by about 3\% and
6\% for Rb$_2$ and Cs$_2$ respectively, while the situation in
Fr$_2$ is less clear as equilibrium distances differ significantly
between Ref.~\cite{lim2005} and the present ones. However the
differences observed among the average polarizabilities stay within
a 10\% range representative of the typical uncertainty of the
experimental values. As we will see in the next section, only the
polarizability values integrated over the wave function of the
lowest vibrational level of the ground state can be directly
compared to the experiment, which may help to discriminate among
theoretical determinations. Only few results have been published for
the lowest triplet state (Table \ref{tab:homo_reqa}), and our values
agree well with them, within a few percent range.

\begin{table}[h]
\begin{tabular} {|c|c|c|c|c|c|c|} \hline
                   &                    &$R_e$ $(a_0)$&$\alpha_{\parallel}$&$\alpha_{\perp}$&$\overline{\alpha}$&$\gamma$ \\ \hline
Li$_2$($a^3\Sigma_u^+$)&
This work                                     &7.88 \cite{linton1989}&700.3&252.2&401.6&448.2\\
&M\'erawa and R\'erat \cite{merawa2001}        &7.88 \cite{linton1989}&698&252&401&446\\
&R\'erat  and Bussery-Honvault \cite{rerat2003}&7.88
\cite{linton1989}&695.8&253.1&400.7&442.7\\ \hline

Na$_2$($a^3\Sigma_u^+$)& This work            &9.62 \cite{li1985}     &495.0&278.2&350.4&216.9\\
&R\'erat  and Bussery-Honvault \cite{rerat2003}&9.62 \cite{li1985}
&487.7&276.9&347.2&210.7\\ \hline

K$_2$($a^3\Sigma_u^+$)&
This work                                     &10.9 \cite{li1990}    &956.4&477.3&637.0&479.1\\
&M\'erawa {\it et al.} \cite{merawa2003}        &10.9 \cite{li1990}
&953.8&476.8&635.8&477.0\\ \hline

Rb$_2$($a^3\Sigma_u^+$)& This work & 11.4
\cite{aymar2006}&1016.4&508.0&677.5&508.4 \\ \hline

Cs$_2$($a^3\Sigma_u^+$)& This work & 11.9
\cite{Li2007}&1322.8&641.2&868.4&681.5\\ \hline
\end{tabular}
\caption {Same as Table \ref{tab:homo_reqX} for the lowest triplet
 state of Li$_2$, Na$_2$, K$_2$, Rb$_2$ and Cs$_2$.}
\label{tab:homo_reqa}
\end{table}

Similarly, we set up a table for polarizabilities of the ground state of heteronuclear alkali pairs (Table \ref{tab:hetero_req}), for which very few other theoretical values are available (except for LiNa): Urban and Sadlej \cite{urban1995} considered the six molecules LiNa, LiK, LiRb, NaK, NaRb, and KRb, in their coupled-cluster approach, while Tarnovsky et al \cite{tarnovsky1993} proposed an estimate of the average polarizability $\overline{\alpha}$ based on their measurements of polarizabilities for homonuclear species (see next section). All determinations are consistent for the lightest species LiNa, while the values of Urban and Sadlej \cite{urban1995} are significantly larger than ours for the heavy species LiRb, NaRb, and KRb.

\begin{table}[h]
\begin{tabular}{|c|c|c|c|c|c|c|} \hline
    &                     &$R_e$ $(a_0)$&$\alpha_{\parallel}$&$\alpha_{\perp}$&$\overline{\alpha}$&$\gamma$ \\ \hline
LiNa&This work                               &5.4518 \cite{engelke1982}&347.6&181.8&237.0&165.8\\
    &Urban and Sadlej \cite{urban1995}       &5.4518 \cite{engelke1982}&352.1&188.8&243.2&163.2\\
    &M\'erawa and Dargelos \cite{merawa1998a}&5.4518 \cite{engelke1982}&351.7&191.5&249.9&160.2\\
    &M\'erawa {\it et al.} \cite{merawa2003a} &5.4518 \cite{engelke1982}&350.6&187.7&242.0&162.9\\
    &Antoine {\it et al.} \cite{antoine1999}  &5.4518 \cite{engelke1982}&352.2&188.9&234.4&163.3\\ \hline
LiK &This work                               &6.268 \cite{engelke1984}&489.7&236.2&320.7&253.5\\
    & Urban and Sadlej \cite{urban1995}      &6.268 \cite{engelke1984}&484.8&246.6&326.0&238.2\\ \hline
LiRb&This work                               &6.5\cite{aymar2005}     &524.3&246.5&339.1&277.8\\
    &  Urban and Sadlej \cite{urban1995}     &6.609                   &558.2&268.7&365.2&289.5\\ \hline
LiCs&This work                               &6.93 \cite{staanum2007}  &597.0&262.5&374.0&334.5\\ \hline
NaK &This work                               &6.61 \cite{krou-adohi1998}&529.2&262.3&351.3&266.9\\
    &  Urban and Sadlej \cite{urban1995}     &6.61 \cite{krou-adohi1998}&537.5&279.6&365.5&257.9\\ \hline
NaRb&This work                               &6.88 \cite{kasahara1996}&572.0&280.3&377.5&291.7 \\
    &Urban and Sadlej \cite{urban1995}       &6.967                   &606.3&303.2&404.2&303.1\\ \hline
NaCs&This work                               &7.27\cite{diemer1984}   &670.7&304.2&426.4&366.5\\ \hline
KRb &This work                               &7.688 \cite{ross1990}   &748.70&382.9&504.8&365.8 \\
    &Urban and Sadlej \cite{urban1995}       &7.786                   &842.4&411.5&555.1&430.9\\ \hline
KCs &This work                               &8.095\cite{ferber2008}    &822.3&425.62&571.1&436.7\\ \hline
RbCs&This work                               &8.366\cite{gustavsson1988}&904.0&492.3&602.8&491.7\\ \hline
\end{tabular}
\caption {Present polarizability values (in a.u.) for the ground
state of heteronuclear alkali dimers, taken at the experimental
(when available) or the theoretical (from Ref.~\cite{aymar2005})
equilibrium distances, and compared to other published theoretical
results.} \label{tab:hetero_req}
\end {table}

\section{Average polarizabilities and their anisotropy for molecular vibrational levels}
\label{sec:exp}

In most experiments the only accessible physical quantity is the
average polarizability of the molecule measured for a given
vibrational level, while the alignment properties is controlled by
the anisotropy of the polarizability. We determine these quantities by averaging their $R$-dependence over the vibrational wave functions $| v \rangle$ of the electronic ground state and of the lowest triplet state of the molecules, i.e.: $\beta_v \equiv \langle v | \beta(R) |v \rangle$, where $\beta$ is one of the quantities $\alpha_{\parallel}$, $\alpha_{\perp}$, $\overline{\alpha}$, $\gamma$. Thus we assumed that the the vibrational motion is not distorted by the polarization of the molecule, following the discussion of ref. \cite{bishop1991}. In any case, the potential curves computed with a non-zero electric field are always found very close to the field-free ones.

\begin{table}[htbp]
\begin{tabular}{|c||c|c|c c|c c||c|c|} \hline
            & \multicolumn{6}{c||}{Singlet ground state} & \multicolumn{2}{c|}{Lowest triplet state} \\
molecule    & $\overline{\alpha} |_{v=0}$ & $\gamma |_{v=0}$  &
$\overline{\alpha}_{\mathsf{max}}$ &  $|v=$  &
$\gamma_{\mathsf{max}}$       &   $|v=$ & $\overline{\alpha}
|_{v=0}$  &  $\gamma |_{v=0}$ \\ \hline
Li$_2$      &   226.8 &  169.6  & 359.0 &   23 &   339.2   &    16  & 399.4  &   438.6\\
Na$_2$      &   259.7 &  179.5  & 357.5 &   40 &   299.3   &    28  & 347.9  &   200.7\\
K$_2$       &   473.0 &  343.7  & 601.2 &   61 &   496.4   &    43  & 636.6  &   476.4\\
Rb$_2$      &   530.6 &  378.8  & 656.1 &   88 &   526.2   &    63  & 685.5  &   493.5\\
Cs$_2$      &   670.3 &  490.8  & 819.2 &  110 &   677.1   &    79
& 865.4  &   667.6 \\ \hline
LiNa        &   236.5 &  166.6  & 339.2 &   37 &   286.3   &    27  &  365.0  &   288.6 \\
LiK         &   318.7 &  250.3  & 474.2 &   38 &   451.5   &    29  &  514.6  &   442.5 \\
LiRb        &   340.4 &  280.4  & 504.5 &   39 &   499.3   &    30  &  535.5  &   429.0 \\
LiCs        &   368.8 &  326.7  & 594.2 &   42 &   654.9   &    33  &  631.0  &   533.7 \\
NaK         &   352.3 &  261.4  & 472.1 &   54 &   407.1   &    38  &  485.8  &   303.5 \\
NaRb        &   375.6 &  288.2  & 501.5 &   61 &   443.1   &    44  &  510.1  &   310.1 \\
NaCs        &   421.9 &  359.4  & 584.1 &   64 &   567.0   &    46  &  589.7  &   351.3 \\
KRb         &   502.0 &  360.3  & 629.1 &   71 &   511.2   &    51  &  661.5  &   484.6 \\
KCs         &   566.0 &  426.7  & 712.0 &   76 &   609.3   &    54  &  749.2  &   564.3 \\
RbCs        &   597.6 &  440.9  & 737.3 &   74 &   609.4   &    53
&  773.2  &   567.5 \\ \hline
\end{tabular}
\caption{Average polarizabilities and anisotropies computed in the
present work for the singlet ground state and lowest triplet states
of all alkali pairs. Values are listed for the lowest vibrational
level ($v=0$), and for the level where the maximum value is reached
for the ground state.} \label{tab:pola_v}
\end {table}

\begin{figure}[f]
\includegraphics[width=0.8\columnwidth]{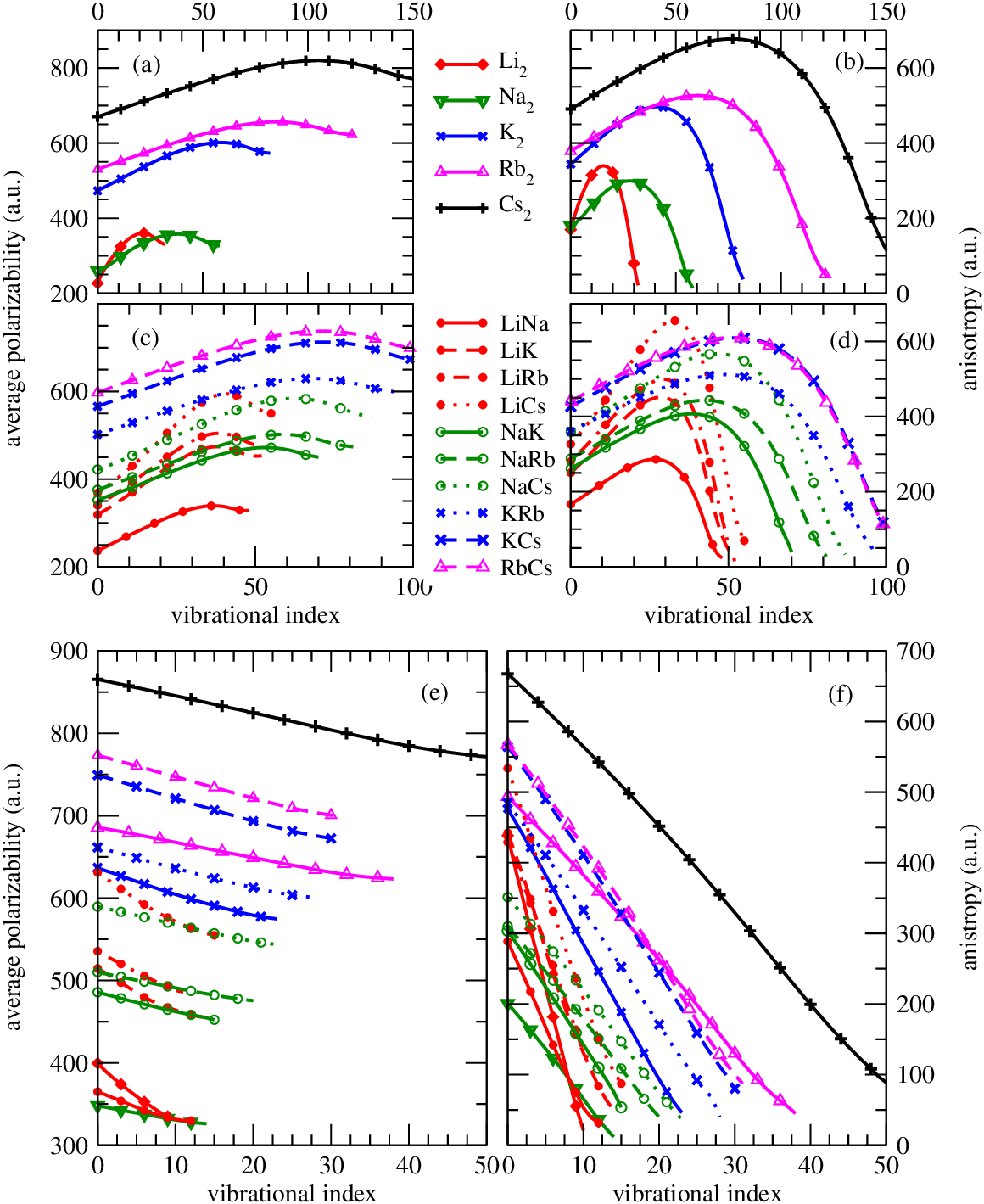}
\caption{\label{fig:avg-ani-pola} (Color on line) Dependence of
average polarizabilities and anisotropies as functions of
vibrational levels: (a)-(d)for the singlet ground state, (e),(f) for
the lowest triplet state.}
\end{figure}

Figure \ref{fig:avg-ani-pola} summarizes the dependence of these
quantities for the two lowest electronic states of all pairs with
the vibrational level, while we extracted in Table \ref{tab:pola_v}
the main relevant features of these variations. As expected from
their monotonic $R$-dependence, the average polarizability of the
lowest triplet states is slowly decreasing down to the sum of atomic
values with increasing vibrational index, while the anisotropy
monotonically drops to zero. In contrast both quantities exhibit a
maximum for a quite high vibrational level of the electronic ground
state, due to the combined influence of the previously mentioned
difference between the equilibrium distance and the position of the
maximum value of the parallel component, and of the increase of the
perpendicular component with the distance.

From the previous section, it is not obvious to discriminate among
the various theoretical determinations, and we examine here if the
reported experimental works can help for this purpose. Only few
experimental works reported values for the average polarizability of
alkali dimers. Molof {\it et al.} \cite{molof1974a} and Tarnovsky
{\it et al.} \cite{tarnovsky1993} measured the deflection of a
thermal molecular beam of alkali dimers in an inhomogeneous electric
field, and both investigated the series of homonuclear species from
Li$_2$ to Cs$_2$. In addition, Tarnovsky {\it et al.}
\cite{tarnovsky1993} also studied the NaK and KCs systems, and
extracted an average static polarizability from their measurement by
subtracting the effect of the permanent dipole moment as calculated
in Ref.~\cite{igel-mann1986}. Using the same technique in their
cluster experiment, Antoine {\it et al.} reported average
polarizabilities for Na$_2$, Li$_2$ and NaLi form a thermal
molecular beam. The drawback of using a thermal beam is that the
deduced values depends on the temperature of the beam, i.e. of the
population of vibrational levels above the lowest one. In contrast
in their experiment on sodium and potassium clusters, Knight {\it et
al} \cite{knight1985} reported on the deflection of a supersonic
molecular beam, in which dimers can safely be considered as being in
the $v=0$ level (i.e. at $T=0$~K). All these values are measured
with a typical 10\% error bar, and they are compiled in Tables
\ref{tab:pola_exp_homo} and \ref{tab:pola_exp_hetero}, together with the present computed average
polarizabilities for the $v=0$ level of the ground state of all
alkali pairs. First we note that the values of Knight {\it et al.}
\cite{knight1985} agree well with those of Tarnovsky {\it et al.}
\cite{tarnovsky1993} extrapolated to $T=0$~K for Li$_2$, Na$_2$ and
K$_2$, which then validates the procedure proposed in
Ref.~\cite{igel-mann1986}. Tarnovsky {\it et al.} suggested that the
low values of Molof {\it et al.} may be due to deviation from the
thermodynamic equilibrium between monomers and dimers in the latter
work. Even if no values extrapolated to $T=0$~K were available for
Rb$_2$ and Cs$_2$, our results still lie within the somewhat large
error bars obtained in the thermal beam of
Ref.~\cite{tarnovsky1993}. This can be understood from the previous
figures as the polarizability smoothly increases for the lowest
vibrational levels. The same conclusion holds for the NaK and KCs
values measured by the same authors, while the two measured values
for LiNa are significantly larger than our prediction. All the other
values for Rb$_2$ and Cs$_2$ and for the heteronuclear species
displayed in Ref.~\cite{tarnovsky1993} are obtained from an
empirical rules involving polarizabilities of homonuclear species
and permanent dipole moments of heteronuclear pairs known at the
time of that work. Their validity is difficult to estimate. However,
as we performed a systematic investigation of all alkali pairs with
similar numerical conditions, we think that such an empirical rules
generally overestimates the average polarizabilities, even if our
values stay within the estimated uncertainty, apart from LiK, LiRb,
and LiCs. In conclusion, the present study show that more
experimental work would be needed, for instance using supersonic
beams of alkali pairs, if they were available.

\begin{table}[h]
\begin{tabular} {|c|c|c|c|c|c|} \hline
 &Li$_2$&Na$_2$&K$_2$ &Rb$_2$&Cs$_2$ \\ \hline
$\overline{\alpha}|_{v=0}$ & 226.8&259.7&473&530.6&670.3 \\ \hline
\cite{tarnovsky1993}(a)&229$\pm$20 (948~K)  &270$\pm$20 (676~K)  &519$\pm$41 (542~K)  &533$\pm$41 (527~K)  &701$\pm$54 (480~K) \\
\cite{tarnovsky1993}(b)&216$\pm$20 (0~K)    &256$\pm$20 (0~K)    &499$\pm$41 (0~K)    &553$\pm$41 (est.)   &675$\pm$54 (est.) \\
\cite{molof1974a} &229$\pm$20 (990~K)&202$\pm$20 (736~K)&411$\pm$34 (569~K)&459$\pm$34 (534~K)&614$\pm$54 \\
\cite{knight1985} &              &263$\pm$20 (0~K)&499$\pm$41 (0~K)&& \\
\cite{antoine1999} &221$\pm$10 (1100~K)&269$\pm$10 (1100~K)&&&\\ \hline
\end{tabular}
\caption{Present average polarizability $\overline{\alpha}|_{v=0}$
(in a.u.) for the ground state of homonuclear alkali dimers, computed for their $v=0$ level. The
values for $\overline{\alpha}$ are compared to the
experimental measurements of Ref.~\cite{molof1974a,knight1985} and
Ref.~\cite{tarnovsky1993} (row (a)) at various temperatures of
their thermal beam, and extrapolated to $T=0$~K following
Ref.~\cite{muller1986} and Ref.~\cite{tarnovsky1993} (row (b)),
i.e. for the $v=0$ level. In this respect, the supersonic beam used
by Knight {\it et al.} \cite{knight1985} is considered to be at
$T=0$~K.} \label{tab:pola_exp_homo}
\end{table}
\begin{table}[h]
\begin{tabular} {|c|c|c|c|c|c|} \hline
& NaK   &KCs   &LiNa  &LiK   &LiRb   \\ \hline
$\overline{\alpha}|_{v=0}$ &352.3&566&236.5&318.7&340.4\\ \hline
\cite{tarnovsky1993}(a)&391$\pm$20 (est.)   &607$\pm$54 (est.) &250$\pm$20 (est.)   &378$\pm$34 (est.) &385$\pm$40 (est.) \\
\cite{tarnovsky1993}(b)&344$\pm$27 (612~K)  &600$\pm$47 (494~K)  & &&\\
\cite{graff1972} &&&270$\pm$30&& \\
\cite{antoine1999}&&&263$\pm$10 (1100~K)&& \\ \hline \hline
&LiCs &NaRb & NaCs  &KRb   &RbCs \\ \hline
$\overline{\alpha}|_{v=0}$&368.8&375.6&421.9&502&597.6 \\ \hline
\cite{tarnovsky1993}&466$\pm$54 (est.)   &398$\pm$40 (est.)   &479$\pm$54 (est.) &526$\pm$40 (est.)   &614$\pm$54 (est.) \\ \hline
\end{tabular}
\caption{Present average polarizability $\overline{\alpha} |_{v=0}$
(in a.u.) ground state of heteronuclear alkali dimers, computed for their $v=0$ level. As for the homonuclears we report the experimental values for NaK and
KCs , as well as the estimates for all the heteronuclear species
 from Ref.~\cite{tarnovsky1993} (see text and Table \ref{tab:pola_exp_homo}, and the experimental
value of Ref.~\cite{antoine1999} for NaLi.} \label{tab:pola_exp_hetero}
\end{table}

\section{Conclusion: prospects for alignment of alkali dimers by external electric fields}
\label{sec:align}

The accurate knowledge of the electric properties of alkali pairs is
particularly relevant in the context of recent developments of
researches in cold molecules. Several authors have recently
addressed the possibility to observe peculiar properties of
ultracold gases of dipolar molecules if they could be partially
aligned or oriented in the presence of external fields
\cite{demille2002,santos2000} to enhance their mutual interaction.
Studies of pendular states using intense laser pulses usually
involve light molecules with static polarizabilities and
anisotropies quite small compared to those of the mixed alkali
pairs. In this context, it is worthwhile to revisit the idea
proposed by Friedrich and Herschbach
\cite{friedrich1999,friedrich1999a,friedrich2000} for aligning polar
molecules by combining an intense laser field and an external
electric field. The interactions between the molecule and the
external fields can be characterized by two dimensionless parameters
$\omega_{or}$ for orientation and $\Delta \omega_{al}$ for
alignment:

\begin{equation}
\omega_{or}=\frac{\mu \varepsilon_S}{B_v}\hspace{1cm}\Delta
\omega_{al}=\frac{\gamma I_L}{2B_v} \label{eq:omega}
\end{equation}

They are related respectively to the interaction potential of the
molecule with permanent dipole moment $\mu$, anisotropy $\gamma$,
and rotational constant $B_v$ in a vibrational level $v$ with an
external static electric field of amplitude $\varepsilon_S$, and
with a laser field of intensity $I_L$. The values of these
orientation and alignment parameters can be conveniently evaluated
with practical units according to: $\omega_{or}=0.0168
\mu(\text{Debye}) \varepsilon_S(\text{kV/cm})/B(\text{cm}^{-1})$,
and $\Delta \omega_{al}=10^{-11}\gamma(\AA^3)
I_L(W/\text{cm}^{-2})/B(\text{cm}^{-1})$.

The pendular hybridization of (polar) molecules by only a static
electric field or only a laser field would require values of these
parameters considerably larger than 1 in order to couple several
rotational states (see also \cite{gonzalez-ferez2004}). Especially
for the orientation of ultracold heteronuclear molecules produced in
a current laser cooling experiment, the application of large static
electric fields is often incompatible with other experimental
requirements such as good optical access to the sample and the fast
switching of high magnetic fields.

Friedrich and Herschbach suggested that the combination of static
and laser fields would result in a double hybridization of the polar
molecules. By plotting electronic wave functions in polar
coordinates, the authors showed that an almost perfect orientation
of a $^1\Sigma$ molecule in the rotational ground state can already
be achieved with typical values of $\omega_{or}=1$ and $\Delta
\omega_{al}=20$.

We display in tables~\ref{tab:efieldX} and \ref{tab:efielda} the
required amplitudes of the fields yielding $\omega_{or}=1$ $\Delta
\omega_{al}=1$ for all alkali pairs, which can easily be scaled to
any field strength depending on the considered experimental
arrangement. Friedrich and Herschbach \cite{friedrich1999a}
displayed such a table for typical values of static fields and laser
pulses intensities, for a series of linear polar molecules like
alkali halides or molecules of atmospheric relevance.

\begin{table}[Htb]
\begin{tabular} {|c|c|c|c|c|c|c|} \hline
&$v_0$&$\gamma$ (a.u.)&$B_v$ ($\times$10$^{-2}$cm$^{-1}$)&$I_{al}$
($\times$10$^8$W/cm$^2$)&$d_v$ (Debye)&$E_{or}$ (kV/cm) \\ \hline
RbCs&0&441&2.90&0.44&-1.237&1.4 \\
KCs &0&427&3.10&0.49&-1.906&1.0 \\
KRb &0&360&3.86&0.72&-0.615&3.7 \\
NaCs&0&359&5.93&1.1&-4.607&0.8 \\
NaRb&0&288&7.11&1.7&-3.306&1.3 \\
NaK &0&261&9.62&2.5&-2.579&2.2 \\
LiCs&0&327&19.4&4.0&-5.523&2.1 \\
LiRb&0&280&22.0&5.3&-4.165&3.1 \\
LiK &0&250&26.1&7.1&-3.565&4.4 \\
LiNa&0&167&38.0&15&-0.566&39.9 \\ \hline
RbCs&77&488&1.65&0.23&-0.906&1.1 \\
KCs &78&483&1.71&0.24&-0.843&1.2 \\
KRb &73&400&2.07&0.35&-0.257&4.8 \\
NaCs&63&483&3.46&0.48&-2.375&0.9 \\
NaRb&61&355&3.82&0.73&-1.558&1.5 \\
NaK &55&321&4.98&1.0&-1.254&2.4 \\
LiCs&41&565&10.8&1.3&-3.051&2.1 \\
LiRb&39&415&11.8&1.9&-1.947&3.6 \\
LiK &39&352&12.7&2.4&-1.205&6.3 \\
LiNa&38&211&16.4&5.3&-0.026&374.9 \\  \hline
\end{tabular}
\caption {Summary of the properties (anisotropy $\gamma$, rotational
constant $B_v$, permanent dipole moment $d_v$) of mixed alkali pairs
relevant for their orientation and alignment induced by external
fields, for the lowest vibrational level of their $X^1\Sigma^+$
ground state. The values for the laser intensity $I_{al}$ of a cw
laser field, and for an external static electric field $E_{or}$
correspond to $\Delta \omega_{al}=1$ and $\omega_{or}=1$
respectively (see text). We also report such properties for the
vibrational level with maximal computed anisotropy.}
\label{tab:efieldX}
\end {table}

The alignment of molecules in pulsed laser fields with typical
intensities of 10$^{12}$\,W/cm$^2$ on timescales of nanoseconds is
an established technique~\cite{stapelfeldt2003}. Alkali dimer ground
states combine large anisotropies, which reduce the required
intensities for alignment, with large permanent dipole moments,
which make permanent orientation of molecules in combined continuous
fields possible. In the following we will describe the experimental
parameters for the permanent orientation of these molecules, using
RbCs as an example. RbCs has a favorable ratio of anisotropy to
rotational constant and has already been produced at ultracold
temperatures in the absolute ground state X$^1\Sigma$ v=0, using
photoassociation followed by a laser-stimulated state transfer
process~\cite{sage2005}. As mentioned above, in order to reach
nearly perfect orientation of the molecule the interactions with the
external fields have to reach at least $\omega_{or}=1$ and $\Delta
\omega_{al}=20$. According to table~\ref{tab:efieldX},
$\omega_{or}=1$ is already reached for a static electric field of
1.4~kV/cm which can be easily realized by a set of largely spaced
electrodes, therefore not reducing the optical access to the sample.
In order to reach $\Delta \omega_{al}=20$, a laser intensity of
roughly 10$^9$\,W/cm$^2$ is necessary. Such intensities can
be reached continuously in a resonator-enhanced dipole
trap~\cite{mosk2001}. Here, Mosk and co-workers have coupled a
1.2\,W Nd:YAG laser beam into an actively stabilized confocal
resonator. The intensity in the anti-nodes of the standing wave
inside the resonator is then given by

\begin{equation} I_0 = 4\times \mathrm{A}
\times \frac{2\,\mathrm{P_L}}{\pi{\omega_0}^2}
\end{equation}

where the factor 4 is due to the coherent addition of fields in a
standing wave, A is the power enhancement factor of the resonator,
P$_L$ the laser power coupled into the cavity and $\omega_0$ the
waist in the focus of the cavity. In Ref.~\cite{mosk2001} a
power-enhancement of nearly A=150 was reported. Using such a setup
with currently available laser powers of the order of 100~W, the
required intensity of 10$^9$\,W/cm$^2$ is reached at a
realistic focus size of $\sim$60$\mu$m. Additionally to the
alignment of the molecules, the anti-nodes of the standing wave
would also trap the molecules via the interaction with the average
polarizability~$\overline\alpha$~\cite{grimm2000}. This leads to the
formation of a stack of pancake-shaped, individually aligned
ensembles at distances of $\lambda/2$ where $\lambda$ is the
wavelength of the alignment laser.

Even the lowest triplet states of alkali dimers appear to be good
candidates for such a combined arrangement, despite their very low
dipole moment. In ultracold LiCs~\cite{kraft2006} for example, values like $\omega_{al}=10$ and $\omega_{or}=0.1$ seem within reach.

\begin{table}[h]
\begin{tabular} {|c|c|c|c|c|c|} \hline
&$\gamma$ (a.u.)&$B_v$ ($\times$10$^{-2}$cm$^{-1}$)&$I_{al}$
($\times$10$^8$W/cm$^2$)&$d_v$ (Debye)&$E_{or}$ (kV/cm) \\ \hline
RbCs&568&1.46&0.17&0.0003&-2887.9 \\
KCs &564&1.53&0.18&-0.013&69.9 \\
KRb &485&1.81&0.25&-0.011&98.1 \\
NaCs&351&2.60&0.50&0.005&-309.3 \\
NaK &303&3.87&0.86&0.008&-287.6 \\
NaRb&310&2.97&0.65&-0.002&883.2 \\
LiCs&534&9.27&1.2&-0.145&38.1 \\
LiRb&429&9.72&1.5&-0.123&47.1 \\
LiK &443&11.5&1.8&-0.113&60.6 \\
LiNa&289&14.1&3.3&-0.068&123.6 \\ \hline
\end{tabular}
\caption {Same as Table \ref{tab:efieldX} for the lowest
$a^3\Sigma^+$ triplet state, all v=0.} \label{tab:efielda}
\end {table}

\begin{acknowledgments}
We are indebted to F. Spiegelman for continuous support all along
our project on electronic structure calculations. Stimulating
discussions with O. Atabek, A. Derevianko and P. Staanum are
gratefully acknowledged. This work also benefited from the expertise
of P. Cahuzac in cluster physics. This work is performed in the
framework of the network "Quantum Dipolar Molecular Gases"
(QuDipMol) of the EUROQUAM program of the European Science
Foundation. JD acknowledges partial support of the French-German
University (http://www.dfh-ufa.org).
\end{acknowledgments}


\begin{thebibliography}{100}

\bibitem{miller1977}
T.~M. Miller and B.~Bederson,
\newblock Adv. At. Mol. Opt. Phys, {\bf 13},\hspace{0.25em}1  (1977).

\bibitem{doyle2004}
J.~Doyle, B.~Friedrich, R.~Krems, and F.~Masnou-Seeuws,
\newblock Eur. Phys. J. D, {\bf 31},\hspace{0.25em}149--164  december 2004.

\bibitem{dulieu2006}
O.~Dulieu, M.~Raoult, and E.~Tiemann,
\newblock J. Phys. B: Atomic Molecular and Optical Physics, {\bf 39}  (2006).

\bibitem{bethlem2003}
H.~L. Bethlem and G.~Meijer,
\newblock Inter. Rev. in Phys. Chem., {\bf 22},\hspace{0.25em}73  (2003).

\bibitem{bochinski2004}
J.~R. Bochinski, Eric~R. Hudson, H.~J. Lewandowski, and Jun Ye,
\newblock Phys. Rev. A, {\bf 70},\hspace{0.25em}043410  (2004).

\bibitem{hudson2006}
E.~R. Hudson, C.~Ticknor, B.~C. Sawyer, C.~A. Taatjes, H.~J. Lewandowski, J.~R.
  Bochinski, J.~L. Bohn, and Jun Ye,
\newblock Phys. Rev. A, {\bf 73},\hspace{0.25em}063404  (2006).

\bibitem{heiner2007}
C.~E. Heiner, D.~Carty, G.~Meijer, and H.~L. Bethlem,
\newblock Nature Physics, {\bf 3},\hspace{0.25em}115  (2007).

\bibitem{junglen2004}
T.~Junglen, T.~Rieger, S.~A. Rangwala, P.W.~H. Pinkse, and G.~Rempe,
\newblock Phys. Rev. Lett., {\bf 92},\hspace{0.25em}223001  (2004).

\bibitem{rieger2006}
T.~Rieger, T.~Junglen, S.~A. Rangwala, G.~Rempe, P.~W.~H. Pinkse, and
  J.~Bulthuis,
\newblock Phys. Rev. A, {\bf 73},\hspace{0.25em}061402  (2006).

\bibitem{gonzalez-ferez2004}
R.~Gonzalez-Ferez and P.~Schmelcher,
\newblock Phys. Rev. A, {\bf 69},\hspace{0.25em}023402  (2004).

\bibitem{vanhaecke2005}
N~Vanhaecke, D~Comparat, and P~Pillet,
\newblock J. Phys. B, {\bf 38},\hspace{0.25em}s409  (2005).

\bibitem{takekoshi1998}
T.~Takekoshi, B.~M. Patterson, and R.~J. Knize,
\newblock Phys. Rev. Lett., {\bf 81},\hspace{0.25em}5105  (1998).

\bibitem{wester2004}
R.~Wester, S.~D.~Kraft, M.~Mudrich, M.~U.~Staudt, J.~Lange,
N.~Vanhaecke, O.~Dulieu, and M.~Weidem\"uller,
\newblock Appl. Phys. B, {\bf 79},\hspace{0.25em}993 (2004)

\bibitem{kraft2005}
S.~D.~Kraft, M.~Mudrich, M.~U.~Staudt, J.~Lange, O.~Dulieu,
R.~Wester, and M.~Weidem\"uller,
\newblock Phys. Rev. A, {\bf 71},\hspace{0.25em}013417 (2005)

\bibitem{mark2007}
M.~Mark, F.~Ferlaino, S.~Knoop, J.~G.~Danzl, T.~Kraemer, C.~Chin,
H.-C.~N\"agerl, and R.~Grimm,
\newblock Phys. Rev. A, {\bf 76},\hspace{0.25em}042514 (2007).

\bibitem{zahzam2006}
N.~Zahzam, T.~Vogt, M.~Mudrich, D.~Comparat, and P.~Pillet,
\newblock Phys. Rev. Lett., {\bf 96},\hspace{0.25em}023202  (2006).

\bibitem{staanum2006}
P.~Staanum, S.~D. Kraft, J.~Lange, R.~Wester, and M.~Weidem\"{u}ller,
\newblock Phys. Rev. Lett., {\bf 96},\hspace{0.25em}023201  (2006).

\bibitem{barker2002}
P.~F. Barker and M.~N. Schneider,
\newblock Phys. Rev. A, {\bf 66},\hspace{0.25em}065402  (2002).

\bibitem{dong2004}
G.~Dong, W.~Lu, and P.~F. Barker,
\newblock Phys. Rev. A, {\bf 69},\hspace{0.25em}013409  (2004).

\bibitem{friedrich1995}
B.~Friedrich and D.~Herschbach,
\newblock Phys. Rev. Lett., {\bf 74},\hspace{0.25em}4623  (1995).

\bibitem{dion1999}
C.~M. Dion, A.~Keller, O.~Atabek, and A.~D. Bandrauk,
\newblock Phys. Rev. A, {\bf 59},\hspace{0.25em}1382  (1999).

\bibitem{sakai1999}
H.~Sakai, C.~P. Safvan, J.~J. Larsen, K.~M. Hilligsoe, K.~Hald, and
  H.~Stapelfeldt,
\newblock J. Chem. Phys., {\bf 110},\hspace{0.25em}10235  (1999).

\bibitem{larsen1999}
J.~J. Larsen, Hi. Sakai, C.~P. Safvan, I.~Wendt-Larsen, and H.~Stapelfeldt,
\newblock J. Chem. Phys., {\bf 111},\hspace{0.25em}7774  (1999).

\bibitem{friedrich1999}
B.~Friedrich and D.~Herschbach,
\newblock J. Chem. Phys., {\bf 111},\hspace{0.25em}6157  (1999).

\bibitem{friedrich2003}
B.~Friedrich, N.~H. Nahler, and U.~Buck,
\newblock J. Mod. Opt., {\bf 50},\hspace{0.25em}2677  (2003).

\bibitem{nahler2003}
N.~H. Nahler, R.~Baumfalk, U.~Buck, Z.Bihary, R.~B. Gerber, and B.~Friedrich,
\newblock J. Chem. Phys., {\bf 119},\hspace{0.25em}224  (2003).

\bibitem{kotochigova2006}
S.~Kotochigova and E.~Tiesinga,
\newblock Phys. Rev. A, {\bf 73},\hspace{0.25em}041405(R)  (2006).

\bibitem{aymar2005}
M.~Aymar and O.~Dulieu,
\newblock J. Chem. Phys., {\bf 122},\hspace{0.25em}204302  (2005).

\bibitem{aymar2007}
M.~Aymar and O.~Dulieu,
\newblock Mol. Phys., {\bf 105},\hspace{0.25em}1733  (2007).

\bibitem{aymar2006}
M~Aymar, O~Dulieu, and F~Spiegelmann,
\newblock J. Phys. B: Atomic Molecular and Optical Physics, {\bf
  39},\hspace{0.25em}S905  (2006).

\bibitem{teachout1971}
R.~R. Teachout and R.~T. Pack,
\newblock AT. Data Nucl. Data Tab., {\bf 3},\hspace{0.25em}195  (1971).

\bibitem{derevianko1999}
A.~Derevianko, W.~R. Johnson, M.~S. Safronova, and J.~F. Babb,
\newblock Phys. Rev. Lett., {\bf 82},\hspace{0.25em}3589  (1999).

\bibitem{safronova1999}
M.~S. Safronova, W.~R. Johnson, and A.~Derevianko,
\newblock Phys. Rev. A, {\bf 60},\hspace{0.25em}4476  (1999).

\bibitem{lim1999}
I.~S. Lim, M. Pernpointner, M. Seth, J.~K. Laerdahl, P.
  Schwerdtfeger, P. Neogrady, and M. Urban,
\newblock Physical Review A (Atomic, Molecular, and Optical Physics), {\bf
  60},\hspace{0.25em}2822--2828  (1999).

\bibitem{magnier2002}
S.~Magnier and M.~Aubert-Fr\'econ,
\newblock J. Quant. Spec. Rad. Transf., {\bf 75},\hspace{0.25em}121  (2002).

\bibitem{molof1974}
R.~W. Molof, H.~L. Schwartz, T.~M. Miller, and B.~Bederson,
\newblock Phys. Rev. A, {\bf 10},\hspace{0.25em}1131  (1974).

\bibitem{chamberlain1963}
G.~E. Chamberlain and J.~C. Zorn,
\newblock Phys. Rev., {\bf 129},\hspace{0.25em}677  (1963).

\bibitem{hall1974}
W.~D. Hall and J.~C. Zorn,
\newblock Phys. Rev. A, {\bf 10},\hspace{0.25em}1141  (1974).

\bibitem{miller1988}
T.~M. Miller and B.~Bederson,
\newblock Adv. At. Mol. Opt. Phys, {\bf 25},\hspace{0.25em}37  (1988).

\bibitem{hunter1991}
L.~R. Hunter, D.~Krause, Jr., D.~J. Berkeland, and M.~G. Boshier,
\newblock Phys. Rev. A, {\bf 44},\hspace{0.25em}6140  (1991).

\bibitem{ekstrom1995}
C.~R. Ekstr\"om, J.~Schmiedmayer, M.~S. Chapman, T.~D. Hammond, and D.~E.
  Pritchard,
\newblock Phys. Rev. A, {\bf 51},\hspace{0.25em}3883  (1995).

\bibitem{miffre2006}
A.~Miffre, M.~Jacquey, M.~B\"uchner, G.~Tr\'enec, and J.~Vigu\'e,
\newblock Phys. Rev. A, {\bf 73},\hspace{0.25em}011603  (2006).

\bibitem{molof1974a}
R.~W. Molof, T.~M. Miller, H.~L. Schwartz, B.~Bederson, and J.~T. Park,
\newblock J. Chem. Phys., {\bf 61},\hspace{0.25em}1816  (1974).

\bibitem{tarnovsky1993}
V.~Tarnovsky, M. Bunimovicz, L. Vuskovi\'{c}, B. Stumpf, and B. Bederson,
\newblock J. Chem. Phys., {\bf 98},\hspace{0.25em}3894  (1993).

\bibitem{antoine1999}
R.~Antoine, D.~Rayane, A.~R. Allouche, M.~Aubert-Fr\'econ, E.~Benichou, F.~W.
  Dalby, Ph. Dugourd, M.~Broyer, and C.~Guet,
\newblock J. Chem. Phys., {\bf 110},\hspace{0.25em}5568  (1999).

\bibitem{graff1972}
J.~Graff, P.J. Dagdigian, and L.~Wharton,
\newblock J. Chem. Phys., {\bf 57},\hspace{0.25em}710  (1972).

\bibitem{knight1985}
W.~D. Knight, K.~Clemenger, W.~A. de~Heer, and W.~A. Saunders,
\newblock Phys. Rev. B, {\bf 31},\hspace{0.25em}2539  (1985).

\bibitem{cohen1965}
H.~D. Cohen and C.~C.~J. Roothaan,
\newblock J. Chem. Phys., {\bf 43},\hspace{0.25em}S34  (1965).

\bibitem{huron1973}
B.~Huron, J.-P. Malrieu, and P.Rancurel,
\newblock J. Chem. Phys., {\bf 58},\hspace{0.25em}5745  (1973).

\bibitem{durand1974}
P.~Durand and J.C. Barthelat,
\newblock Chem. Phys. Lett., {\bf 27},\hspace{0.25em}191  (1974).

\bibitem{durand1975}
P.~Durand and J.C. Barthelat,
\newblock Theor. chim. Acta, {\bf 38},\hspace{0.25em}283  (1975).

\bibitem{muller1984}
W.~M\"uller and W.~Meyer,
\newblock J. Chem. Phys., {\bf 80},\hspace{0.25em}3311  (1984).

\bibitem{foucrault1992}
M.~Foucrault, Ph. Milli\'e, and J.P. Daudey,
\newblock J. Chem. Phys., {\bf 96}(2),\hspace{0.25em}1257  (1992).

\bibitem{kello1993}
V.~Kell\"o, A.~J. Sadlej, and K.~Faegri Jr.,
\newblock Phys. Rev. A, {\bf 47},\hspace{0.25em}1715  (1993).

\bibitem{muller1986}
W.~M\"uller and W.~Meyer,
\newblock J. Chem. Phys., {\bf 85},\hspace{0.25em}953  (1986).

\bibitem{rerat2003}
M.~R\'erat and B.~Bussery-Honvault,
\newblock Mol. Phys., {\bf 101},\hspace{0.25em}373  (2003).

\bibitem{merawa2003}
M.~M\'erawa, M.~R\'erat, and B.~Bussery-Honvault,
\newblock J. Molec. Struct., {\bf 633},\hspace{0.25em}137  (2003).

\bibitem{heijmen1996}
T.~G.~A. Heijmen, R.~Moszynski, P.~E.~S. Wormer, and A.~Van~Der Avoird,
\newblock Mol. Phys., {\bf 89},\hspace{0.25em}81  (1996).

\bibitem{merawa2003a}
M.~M\'erawa, D.~B\'egu\'e, and A.~Dargelos,
\newblock Chem. Phys. Lett., {\bf 372},\hspace{0.25em}529  (2003).

\bibitem{politzer2002}
P.~Politzer, P.~Jin, and J.~S. Murray,
\newblock J. Chem. Phys., {\bf 117},\hspace{0.25em}8197  (2002).

\bibitem{laidig1990}
K.~E. Laidig and R.~F.~W. Bader,
\newblock J. Chem. Phys., {\bf 93},\hspace{0.25em}7213  (1990).

\bibitem{brinck1993}
T.~Brinck, J.~S. Murray, and P.~Politzer,
\newblock J. Chem. Phys., {\bf 98},\hspace{0.25em}4305  (1993).

\bibitem{gough1989}
K.~M. Gough,
\newblock J. Chem. Phys., {\bf 91},\hspace{0.25em}2424  (1989).

\bibitem{chandrakumar2004}
K.~R.~S. Chandrakumar, T.~K. Ghanty, and S.~K. Ghosh,
\newblock J. Chem. Phys., {\bf 120},\hspace{0.25em}6487  (2004).

\bibitem{apell2002}
S.~P. \"Apell, J.~R. Sabin, S.~B. Trickey, and J.~Oddershede,
\newblock Intern. J. Quantum Chem., {\bf 86},\hspace{0.25em}35  (2002).

\bibitem{ambjornsson2003}
T.~Ambj\"ornsson and S.~P. \"Apell,
\newblock Phys. Rev. E, {\bf 67},\hspace{0.25em}031917  (2003).

\bibitem{Kornyushin2004}
Y.~Kornyushin,
\newblock Science of Sintering, {\bf 36},\hspace{0.25em}43  (2004).

\bibitem{bishop1984}
D.~M. Bishop and C.~Pouchan,
\newblock J. Chem. Phys., {\bf 80},\hspace{0.25em}789  (1984).

\bibitem{bishop1985a}
D.~M. Bishop, M.~Chaillet, C.~Larrieu, and C.~Pouchan,
\newblock Phys. Rev. A, {\bf 31},\hspace{0.25em}2785  (1985).

\bibitem{spelsberg1993}
D.~Spelsberg, T.~Lorenz, and W.~Meyer,
\newblock J. Chem. Phys., {\bf 99},\hspace{0.25em}7845  (1993).

\bibitem{moullet1989}
I.~Moullet, J.~L. Martins, F.~Reuse, and J.~Buttet,
\newblock Z. Phys.D, {\bf 12},\hspace{0.25em}353  (1989).

\bibitem{merawa1998a}
M.~M\'erawa and A.~Dargelos,
\newblock J. Chim. Phys. Phys. Chim. Bio., {\bf 95},\hspace{0.25em}1711
  (1998).

\bibitem{maroulis1988}
G.~Maroulis,
\newblock Mol. Phys., {\bf 63},\hspace{0.25em}299  (1988).

\bibitem{maroulis2004}
G.~Maroulis,
\newblock J. Chem. Phys., {\bf 121},\hspace{0.25em}10519  (2004).

\bibitem{urban1995}
M.~Urban and A.~J. Sadlej,
\newblock J. Chem. Phys., {\bf 103},\hspace{0.25em}9692  (1995).

\bibitem{lim2005}
I.~S. Lim, P. Schwerdtfeger, T. S\"ohnel, and H. Stoll,
\newblock J. Chem. Phys., {\bf 122},\hspace{0.25em}134307  (2005).

\bibitem{barakat1986}
B.~Barakat, R.~Bacis, F.~Carrot, S.~Churassy, P.~Crozet, F.~Martin, and
  J.~Verg\`es,
\newblock Chem. Phys., {\bf 102},\hspace{0.25em}215  (1986).

\bibitem{merawa2001}
M.~M\'erawa and M.~R\'erat,
\newblock Eur. Phys. J. D, {\bf 17},\hspace{0.25em}329  (2001).

\bibitem{babaky1988}
O.~Babaky and K.~Hussein,
\newblock Can. J. Phys., {\bf 67},\hspace{0.25em}912  (1988).

\bibitem{amiot1991}
C.~Amiot,
\newblock J. Molec. Spect., {\bf 147},\hspace{0.25em}370  (1991).

\bibitem{engelke1984a}
F.~Engelke, H.~Hage, and U.~Schuhle,
\newblock Chem. Phys. Lett., {\bf 106},\hspace{0.25em}535  (1984).

\bibitem{seto2000}
J.~Y. Seto, R. J.~Le Roy, J. Verg\`es, and C. Amiot,
\newblock J. of Chem. Phys., {\bf 113},\hspace{0.25em}3067 (2000).

\bibitem{caldwell1980}
C.~D. Caldwell, F.~Engelke, and H.~Hage,
\newblock Chem. Phys., {\bf 54},\hspace{0.25em}21  (1980).

\bibitem{amiot2002a}
C.~Amiot and O.~Dulieu,
\newblock J. Chem. Phys., {\bf 117},\hspace{0.25em}5155  (2002).

\bibitem{linton1989}
C.~Linton, T.~L. Murphy, F.~Martin, R.~Bacis, and J.~Verg\`es,
\newblock J. Chem. Phys., {\bf 91},\hspace{0.25em}6036  (1989).

\bibitem{li1985}
L.~Li, S.~F. Rice, and R.~W. Field,
\newblock J. Chem. Phys., {\bf 82},\hspace{0.25em}1178  (1985).

\bibitem{li1990}
L.~Li, A.~M. Lyyra, W.~T. Luh, and W.~C. Stwalley,
\newblock J. Chem. Phys., {\bf 93},\hspace{0.25em}8452  (1990).

\bibitem{Li2007}
D.~Li, F.~Xie, L.~Li, S.~Magnier, V.B. Sovkov, and V.S. Ivanov,
\newblock Chem. Phys. Lett., {\bf 441},\hspace{0.25em}39  (2007).

\bibitem{engelke1982}
F.~Engelke, G.~Ennen, and K.~H. Meiwes,
\newblock Chem. Phys., {\bf 66},\hspace{0.25em}391  (1982).

\bibitem{engelke1984}
F.~Engelke, H.~Hage, and U.~Sprick,
\newblock Chem. Phys., {\bf 88},\hspace{0.25em}443  (1984).

\bibitem{staanum2007}
P~Staanum, A.~Pashov, H.~Kn\"ockel, and E.~Tiemann,
\newblock Phys. Rev. A, {\bf 75},\hspace{0.25em}042513  (2007).

\bibitem{krou-adohi1998}
A.~Krou-Adohi and S.~Giraud-Cotton,
\newblock J. Molec. Spect., {\bf 190},\hspace{0.25em}171  (1998).

\bibitem{kasahara1996}
S.~Kasahara, T.~Ebi, M.Tanimura, H.~Ikoma, K.~Matsubara, M.~Baba, and H.~Kato,
\newblock J. Chem. Phys., {\bf 105},\hspace{0.25em}1341 (1996).

\bibitem{diemer1984}
U.~Diemer, H.~Weickenmeier, M.~Wahl, and W.~Demtr\"oder,
\newblock Chem. Phys. Lett., {\bf 104},\hspace{0.25em}489  (1984).

\bibitem{ross1990}
A.~J. Ross, C.~Effantin, P.~Crozet, and E.~Boursey,
\newblock J. Phys. B, {\bf 23},\hspace{0.25em}L247  (1990).

\bibitem{ferber2008}
R.~Ferber, I.~Klincare, O.~Nikolayeva, M.~Tamanis, H.~Kn\"ockel, E.~Tiemann, and A.~Pashov
\newblock Phys. Rev. A, {\it in press}

\bibitem{gustavsson1988}
T.~Gustavsson, C.~Amiot, and J.~Verg\`es,
\newblock Chem. Phys. Lett., {\bf 143},\hspace{0.25em}101  (1988).

\bibitem{bishop1991}
D. M.~Bishop and B.~Kirtman
\newblock J. Chem. Phys., {\bf 95},\hspace{0.25em}2646  (1991).

\bibitem{igel-mann1986}
G.~Igel-Mann, U.~Wedig, P.~Fuentealba, and H.~Stoll,
\newblock J. Chem. Phys., {\bf 84},\hspace{0.25em}5007  (1986).

\bibitem{demille2002}
D.~DeMille,
\newblock Phys. Rev. Lett., {\bf 88},\hspace{0.25em}067901  (2002).

\bibitem{santos2000}
L.~Santos, G.~V. Shlyapnikov, P.~Zoller, and M.~Lewenstein,
\newblock Phys. Rev. Lett., {\bf 85},\hspace{0.25em}1791  (2000).

\bibitem{friedrich1999a}
B.~Friedrich and D.~Herschbach,
\newblock J. Phys. Chem. A, {\bf 103},\hspace{0.25em}10280  (1999).

\bibitem{friedrich2000}
B.~Friedrich,
\newblock Phys. Rev. A, {\bf 61},\hspace{0.25em}025403  (2000).

\bibitem{stapelfeldt2003}
H.~Stapelfeldt and T.~Seideman
\newblock  Rev. Mod. Phys., {\bf 75},\hspace{0.25em}543  (2003).

\bibitem{sage2005}
J.~M.~Sage, S.~Sainis, T.~Bergeman, and D.~DeMille,
\newblock  Phys. Rev. Lett., {\bf 94},\hspace{0.25em}203001 (2005).

\bibitem{mosk2001}
A.~Mosk, S. Jochim, H. Moritz, Th. Els\"asser, M. Weidem\"uller, and R. Grimm
\newblock  Opt. Lett., {\bf 23},\hspace{0.25em}1837 (2001).

\bibitem{grimm2000}
R.~Grimm, M.~Weidem{\"u}ller, and Y.~B. Ovchinnikov,
\newblock  Adv. At. Mol. Opt. Phys., {\bf 42},\hspace{0.25em}95 (2000).

\bibitem{kraft2006}
S.~D.~Kraft, P.~Staanum, J.~Lange, L.~Vogel, R.~Wester, and
M.~Weidem\"uller,
\newblock  J. Phys. B,{\bf 39}, S99 (2006).

\end{thebibliography}

\newpage

\end{document}